\documentclass[a4paper,11pt]{article}
\pdfoutput=1 

\usepackage{jcappub} 

\usepackage[T1]{fontenc} 
\usepackage{natbib}
\usepackage{subfigure}
\usepackage[nottoc]{tocbibind}
\usepackage{booktabs}
\usepackage{multirow}
\usepackage{enumitem}
\usepackage{longtable}
\usepackage{xcolor}
\usepackage{multirow}
\usepackage{soul}

\title{\boldmath Clustering of Gravitational Wave and Supernovae events: \\ a multitracer analysis \\ in Luminosity Distance Space} 

\author[a,b]{S. Libanore,}
\author[a,c,g]{M.C. Artale,}
\author[d]{D. Karagiannis,}
\author[a,b]{M. Liguori,}
\author[a,b]{\\N. Bartolo,}
\author[a,b]{Y. Bouffanais,}
\author[a,b,e]{M. Mapelli,}
\author[a,b,e,f]{S. Matarrese}

\affiliation[a]{Dipartimento di Fisica e Astronomia G. Galilei, Universit\`{a} degli Studi di Padova \\ Via Marzolo 8, 35131 Padova, Italy}
\affiliation[b]{INFN, Sezione di Padova, Via Marzolo 8, I$-$35131, Padova, Italy}
\affiliation[c]{Institut f$\ddot{\text{u}}$r Astro- und Teilchenphysik, Universit$\ddot{\text{a}}$t Innsbruck\\ Technikerstrasse 25/8, 6020 Innsbruck, Austria}
\affiliation[d]{Department of Physics and Astronomy, University of the Western Cape\\ Cape Town 7535, South Africa}
\affiliation[e]{INAF, Osservatorio Astronomico di Padova\\ Vicolo dell'Osservatorio 5, I$-$35122, Padova, Italy}
\affiliation[f]{Gran Sasso Science Institute (GSSI), Viale F. Crispi 7, I$-$67100, L'Aquila, Italy}
\affiliation[g]{Department of Physics and Astronomy, Purdue University, 525 Northwestern Avenue, West Lafayette, IN 47907, USA}

\emailAdd{sarah.libanore@phd.unipd.it}

\abstract{We study the clustering of Gravitational Wave (GW) merger events and Supernovae IA (SN), as cosmic tracers in Luminosity Distance Space. We modify the publicly available \texttt{CAMB} code to numerically evaluate auto- and cross- power spectra for the different sources, including Luminosity Distance Space distortion effects generated by peculiar velocities and lensing convergence. We perform a multitracer Fisher analysis to forecast expected constraints on cosmological and GW bias coefficients, using outputs from hydrodynamical N-body simulations to determine the bias fiducial model and considering future observations from the Vera Rubin Observatory and Einstein Telescope (ET), both single and in a 3 detector network configuration.  We find that adding SN to the GW merger dataset considerably improves the forecast, mostly by breaking significant parameter degeneracies, with final constraints comparable to those obtainable from a {\it Euclid}-like survey. GW merger bias is forecasted to be detectable with good significance even in the single ET case.}

\begin{document}
\maketitle
\flushbottom

\newpage 

\section{Introduction}

The last six years have witnessed the first three observing runs of Advanced LIGO and Virgo, leading to the detection of more than 50 gravitational wave (GW) events  \cite{abbottO3a,abbottGWTC2.1,abbott_2021} from compact binary mergers. This sample will conspicuously grow in the next two observing runs \cite{Abbott_2020}. Furthermore, third generation GW detectors, such as Einstein Telescope\footnote{\texttt{http://www.et-gw.eu}} (ET), will lead to a dramatic increase of the detection rate, observing $\sim{10^5}$ GW events per year with accurate luminosity distance determination \cite{maggiore2020}. The sky localization precision will be in a range between tenths to a hundred deg$^2$ for a single detector and it will significantly increase using networks of third generation detectors (both ET-like or Cosmic Explorer\footnote{\texttt{https://cosmicexplorer.org/}}-like). 
This scenario will allow us to perform statistical studies of the clustering properties of GW mergers. Since these are tracers of the Large Scale Structure (LSS) of the Universe, this will enable us to constrain both cosmological and merger bias parameters, both cross-correlating with galaxy maps \cite{scelfo_2018, muk_2021_bias,CAlore_2020,Ca_as_Herrera_2021} and by using GW surveys alone \cite{Namikawa_2016,zhang2018,Libanore_2021,vijaykumar2020probing}.

For analogous studies of 3D galaxy clustering, we use redshifts as distance indicators; however, for GW events we directly measure luminosity distances, not redshifts. If we want to identify the redshifts of GW sources, we need to rely on either electromagnetic counterparts or statistical methods, based on cross-correlations with theoretical models or other LSS tracers. These approaches present some drawbacks: both rely on extra-observations or external datasets and are difficult or impossible to apply for high redshift sources; moreover, the statistical analysis can depend on specific assumptions. The natural alternative to redshift identification, which we consider in this work, is that of directly using luminosity distance as our distance indicator in the 3D clustering analysis, as originally proposed in \cite{Namikawa_2016,zhang2018}. In this way, no external datasets or assumptions are needed and all mergers, including very high redshift ones, can be included in the analysis.

In a previous work \cite{Libanore_2021}, we produced cosmological and merger bias parameter forecasts, considering third generation interferometers and more futuristic scenarios. We underlined that the main advantages of this kind of survey are the large volumes probed and the capability of constraining the merger bias at high statistical significance, which can provide interesting information about the physical nature and properties of mergers themselves. In this paper we significantly expand our previous study, in three main directions. 

Firstly, we include new tracers. In our previous analysis, we considered only GW events produced by Double Neutron Star (DNS) and Double Black Hole (DBH) mergers; the same procedure can be used to analyse Black Hole-Neutron Star mergers (BHNS) as well. Moreover, another type of observable for which a Luminosity Distance Space (LDS) clustering analysis is natural exists, namely Supernovae IA (SN). Here, we therefore include SN in the analysis.

Secondly, we improve our treatment of Luminosity Distance Space Distortions (LDSD). Differently from redshift space distortions (RSD), LDSD display a leading order contribution coming from the Line-of-Sight (LoS) derivative of the lensing convergence. In \cite{Libanore_2021}, following an heuristic argument presented in \cite{zhang2018}, we neglected this contribution in our power spectrum calculation. Lensing of GW is an active field of study (see e.g., \cite{Cremonese_2021} for a treatment of lensing effects in the GW signal and e.g. \cite{Mukherjee_2020_lensing} for a weak lensing analysis). As for LDSD, the lensing term was explicitly derived and discussed in \cite{Namikawa_2021}, where the resulting merger power spectrum was evaluated using a Limber approximation. Following that calculation, here we implement a modified version of the \texttt{CAMB} code, which evaluates this term in full sky, so that we can propagate and assess its impact on the final parameter forecasts. We find that the lensing LDSD contribution only affects the large scales and has a very small impact on the final parameter forecasts. These findings are in line with both the power spectrum computation\footnote{Note that for this analysis we need to consider only the monopole term. As shown in \cite{Namikawa_2021}, the lensing contribution has a larger effect and a distinctive signature on higher order multipoles.} of \cite{Namikawa_2021} and the general argument raised in \cite{zhang2018}.

Finally, we go beyond the single tracer approach of our previous work and develop a multitracer analysis. We show that SN play a significant role in breaking degeneracies arising in a GW-only analysis, thus leading to important improvements in the expected constraints.
As in our previous work, we obtain our fiducial values for merger bias parameters by building an Halo Occupation Distribution (HOD), based on full hydrodynamical simulations \cite{Artale2018,Artale2019}; the fiducial model for SN is instead based on the literature \cite{Mukherjee_2020_lensing}. We also study how the inclusion of a prior on one of such parameters affects the results.
Our full multitracer analysis is performed assuming that GW observations are made using Einstein Telescope either in single (ET) or in three detector network configuration (ET$\times$3). For Supernovae IA instead, the Vera Rubin Observatory\footnote{\texttt{https://www.lsst.org}} (VRO), previously known as Large Synoptic Survey Telescope (LSST), is considered. 

This paper is structured as follows: in sec. \ref{sec:tracer} we describe the number distributions and bias models of the tracers (mergers and SN) used in the analysis. In sec. \ref{sec:LDS} we present the observational effects that must be taken into account when mapping the tracers in LDS, namely the Luminosity Distance Space Distortions and the lensing (other effects are neglected in this work being subdominant and still not well described at the theoretical level). Sec. \ref{sec:method} presents the formalism we used to perform the single and multitracer analysis, while sec. \ref{sec:results} describes the constraints we obtained both on cosmological and bias parameters.

\section{Tracer distributions and bias}\label{sec:tracer}

Working in Luminosity Distance Space can present significant advantages when studying the spatial distribution of sources such as GW produced by compact binary mergers and SN. 
On the one hand, binary coalescences and merger signals are very well described (see e.g. \cite{LIGO2016}): from their amplitude and frequency variation it is possible to extract a full set of intrinsic and observational parameters, including their luminosity distance $D_L$ \cite{Veitch_2015,yang2019}. 
On the other hand, the explosion mechanism which produces the SN determines the light curve shape (see e.g. \cite{Khokhlov1993}) and the absolute magnitude of the peak, allowing us to estimate $D_L$ \cite{Jensen2004} after a proper calibration procedure.
Since we assume General Relativity in our work, the luminosity distance $D_L$ represents the same kind of measurement for both merger and SN observations. In some scenarios involving dark energy or modified gravity, however, the luminosity distance extracted from the gravitational wave signal $D_L^{GW}$ turns out to be different from the electromagnetic one $D_L^{EM}$, both in the background (see e.g. \cite{Belgacem_2018,Belgacem2_2018}) and when considering fluctuations (compare with \cite{Garoffolo_2020,Dalang_2020}), making the radial coordinates $D_L^{GW}$ and $D_L^{EM}$ different. 
In these scenarios, our analysis would be self-consistent except for the multitracer case with combined mergers and SN (sec. \ref{sec:results_multi}). As shown by \cite{Garoffolo_2021}, the combination of the two tracers would still be possible using other techniques and it should allow constraining the dark energy or modified gravity models themselves.

To study the constraining power of future LDS surveys, first and foremost the distribution and clustering of the tracers have to be modelled.
In sec. \ref{sec:GW}, we model the distribution and clustering properties of GW referring to compact binary mergers by means of numerical simulations, suitably processed to account for observational selection effects. In sec. \ref{sec:SN}, we instead model the SN distribution, by taking into account the event rate and completeness of the observations performed. 

Despite the analysis is performed in LDS, along the paper the redshift notation is frequently used, since it is more familiar to the reader and more similar to the one used when studying LSS. The consistency of the analysis is preserved since the redshift is always computed from luminosity distance assuming the {\it Planck 2018} \cite{Planck2018} Cosmology.

\subsection{Gravitational Wave mergers}\label{sec:GW}

In our previous work \cite{Libanore_2021}, we explicitly derived the number distribution and bias of DBH and DNS mergers; here, we also consider analogously BHNS systems.  
In doing so, we rely on the simulations discussed in \cite{Artale2018,Artale2019} to get the probability distribution of the mergers, as a function of their redshift $z$ and of the host galaxy stellar mass $M_*$ and star formation rate $SFR$. The population-synthesis simulations and their coupling with the cosmological simulation are the same as we described in \cite{Libanore_2021}. 
We define the total number of mergers in the simulations as:
\begin{equation}
N_m^{SIM}(z) = \sum_{i}\sum_{j}\bigl<N^{SIM}_m(z)|M_*^{i},SFR^j\bigr>\ ,    
\end{equation}
where $m$ indicates DBH, BHNS or DNS, while $i$, $j$ indicate respectively the $M_*$ and $SFR$ bins used in the simulation. The brackets $\bigl<...\bigr>$ indicates that we are computing the expected value of the quantity they contain i.e., the probability distribution of $N_m^{SIM}(z)$ conditioned over the values of $M_*^i,\ SFR^j$ ($z$, instead, is considered as a parameter). Selection effects due to the ET Signal-to-Noise Ratio (SNR) are taken into account inside the simulations; the detection rate is then directly computed by converting the time intervals from the source to the observer rest frame, through:
\begin{equation}\label{eq:N_merger}
N_m(z)\ =\ T^{OBS}_m\ \frac{N_m^{SIM}(z)}{T^{SIM}(z)}\ \frac{1}{1+z}\ .
\end{equation} 
$T^{SIM}(z)$ represents the width of the redshift bins used for the simulation snapshots $[z-\delta z, z+\delta z]$ expressed in Gyr, while $T^{OBS}_m$ is the duration of the observational campaign, which we assume to be 3 years long. 

The observed number density per unit redshift and solid angle can be expressed as: 
\begin{equation}\label{eq:dNdzdOmega}
\frac{d^2N_m}{dzd\Omega} = N_m(z)\frac{c}{\ell\ H(z)}\biggl(\frac{D_L(z)}{\ell\ (1+z)^2}\biggr)^{2}\ ,
\end{equation}
where $\ell = 25$Mpc is the overall size of the simulation box, $H(z)$ is the Hubble factor and $c$ is the speed of light. 
Here, we adopt the same procedure used in \cite{Libanore_2021} and interpolate the source number density measured from simulations with a skewed Gaussian; the resulting distributions are shown in fig. \ref{fig:allsources}. 

The probability distribution $\bigl<N^{SIM}_m(z)|M_*^{i},SFR^j\bigr>$ can be used also to compute the bias of the mergers using an HOD based approach.
We follow the two-step procedure we discussed in detail in \cite{Libanore_2021} and start by computing the bias of host galaxies in each $M_*$ and $SFR$ bin as: 
\begin{equation}\label{eq:biasgal_m}
b_g(z,M_*,SFR) = \int_{M_h^{min,(M_*,SFR)}}^{+\infty} dM_h \ n_h(z,M_h) \ b_h(z,M_h)\frac{\bigl<N_g(M_*,SFR)|M_h\bigr>}{n_g(z,M_*,SFR)}\ ,
\end{equation}
where $n_h(z,M_h)$, $b_h(z,M_h)$ are the halo mass function and the bias factor \cite{Tinker2008} respectively, while $M_h^{min,(M_*,SFR)}$ represents the minimum mass required for an halo to form a galaxy with a given stellar mass $M_*$ and star formation rate $SFR$. Moreover,  $\bigl<N_g(M_*,SFR)|M_h\bigr>$ is the galaxy Halo Occupation Distribution (HOD) in the \textsc{eagle} simulation \cite{Schaye2015}, defined as the number of galaxies with stellar mass $M_*$ and star formation rate $SFR$, within a dark matter halo of mass $M_h$. The galaxy mean number density, $n_g(z,M_*,SFR)$, is defined as:
\begin{equation}\label{eq:gal_den}
n_g(z,M_*,SFR)= \int_{M_h^{min,(*,SFR)}}^{+\infty} dM_h\ n_h(z,M_h)\ \bigl<N_g(M_*,SFR)|M_h\bigr> \ .
\end{equation}
The merger bias can then be computed in the second step, through:
\begin{equation}\label{eq:biasmerger}
b_m(z) = \int_{M_*^{min}}^{M_*^{max}} dM_*\ \int_{SFR^{min}}^{SFR^{max}} dSFR\ n_g(z,M_*,SFR)\ b_g(z,M_*,SFR)\frac{\bigl<N_m(z)|M_*,SFR\bigr>}{n_m(z)} \ .
\end{equation}

\noindent
In the previous equation, $n_m(z)$ represents the merger mean number density, which is computed in analogy to eq. (\ref{eq:gal_den}) through:
\begin{equation}
n_m(z)= \int_{M_*^{min}}^{M_*^{max}}\int_{SFR^{min}}^{SFR^{max}} dM_*\ dSFR\ n_g(z,M_*,SFR)\ \bigl<N_m^{SIM}(z)|M_*,SFR\bigr> \ .
\end{equation}

\noindent
Results obtained in \cite{Libanore_2021} show that the bias for both DBH and DNS can be modelled as:
\begin{equation}\label{eq:bias}
b_m(z) = A_m(z+B_m)^{P_m} \ , 
\end{equation}
where the slope $P_m$ is well described by a linear behaviour i.e., $P_m = 1$. $A_m$ and $B_m$ fiducial values are extracted from the simulations, obtaining for the different kinds of mergers the following values: $A_{DBH} = 0.7$, $B_{DBH} = 2.68$, $A_{DNS} = 0.76$, $B_{DNS} = 2.46$. Here, we verify that eq. (\ref{eq:bias}) well describes also the BHNS bias, once that the fiducial values $A_{BHNS} = 0.71$, $B_{BHNS} = 2.63, P_{BHNS} = 1$ are used. These results are summarized in fig. \ref{fig:allbias}.

\subsection{Supernovae IA}\label{sec:SN}

\begin{figure}[ht!]
    \centering
    \includegraphics[scale=0.4]{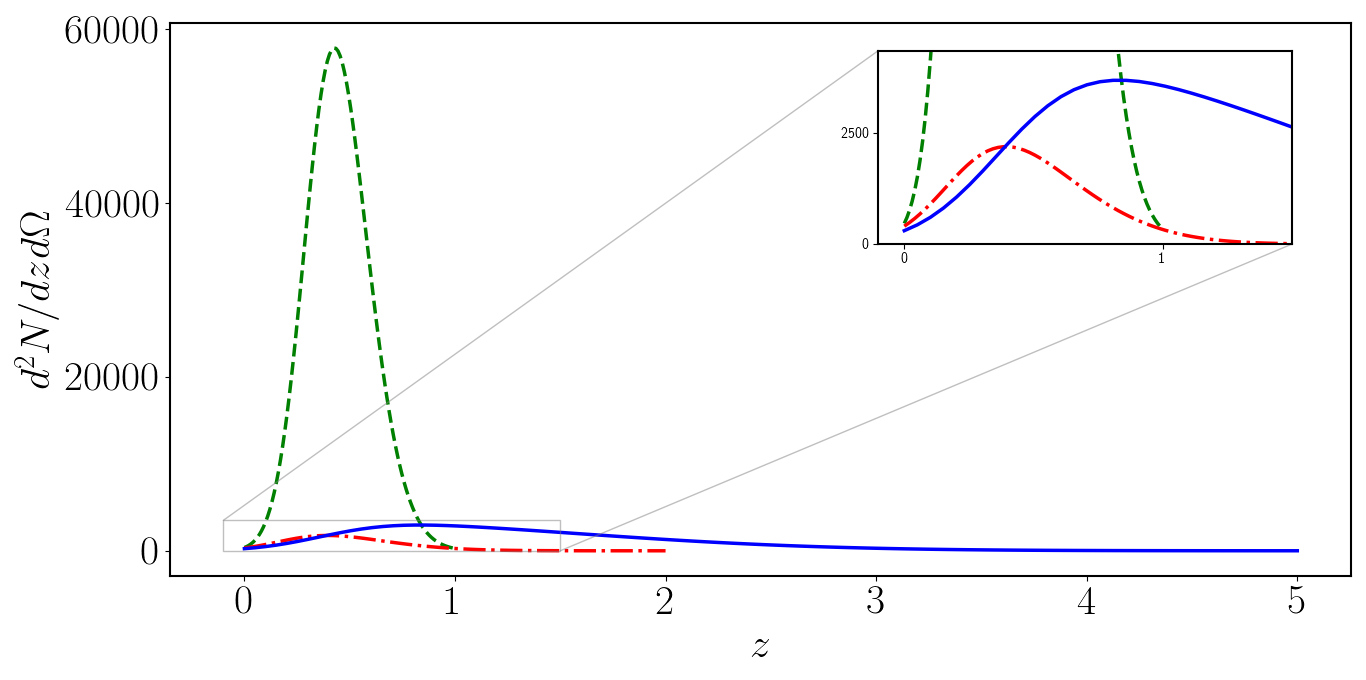}
    \caption{$d^2N/dzd\Omega$ for the sources considered in the analysis: SN observed by the Vera Rubin Observatory in the \textit{Status Quo} configuration (green dashed line, see sec. \ref{sec:SN}), DNS (red dotted-dashed line, see sec. \ref{sec:GW}) and DBH observed by ET (blue line, see sec. \ref{sec:GW}). $z_{max}$ per each source depends on the detector horizon (see tab. \ref{tab:survey}). BHNS observed by ET are not showed for clarity; their model is in between DNS and DBH.}
    \label{fig:allsources}
\end{figure}

To model the SN number distribution, we combine the SN rate provided in \cite{LSST20019} with the {\it Status Quo} completeness that the work by \cite{Garcia2020} describes for the Vera Rubin Observatory. We assume a $T^{OBS}_{SN} = 5$yr observational run. 
The rate in the source rest frame is modelled as:
\begin{equation}
    r_{IA} = 2.6\cdot 10^{-5}(1+z)^{2.5} \ h^3_{70}\ \text{SN yr}^{-1}\text{Mpc}^{-3} \ ,
\end{equation}
where $h_{70}$ is the rescaling of the Hubble parameter determined by the fiducial Cosmology. The conversion to the observer rest frame is performed using the same factor as in eq. (\ref{eq:N_merger}). 
The completeness of the survey, i.e., the fraction of events that are detected, is modelled as: 
\begin{equation}
    \mathcal{C} = \gamma_{SN} \biggl(\frac{z}{\alpha_{SN}}\biggr)^{\beta_{SN}-1}e^{(-z/\alpha_{SN})^{\beta_{SN}}}\ ,
\end{equation}
where $\alpha_{SN}$, $\beta_{SN}$, $\gamma_{SN}$ are chosen to reproduce the completeness trend in \cite{Garcia2020}.
A comoving box having size $\ell = 1$Mpc is considered to compute the observed number of SN events as:
\begin{equation}\label{eq:N_SN}
    N_{SN}(z) = T^{OBS}_{SN}\mathcal{C}\ \ell^{3}\ r_{IA}\ . 
\end{equation}
The distribution we obtain in this way is shown in fig. \ref{fig:allsources}; the total number of events observed is $\sim 10^{4.86}$, which is slightly more pessimistic than what both \cite{Howlett2017} and \cite{Garcia2020} report.

\begin{figure}[ht!]
    \centering
    \includegraphics[scale=0.4]{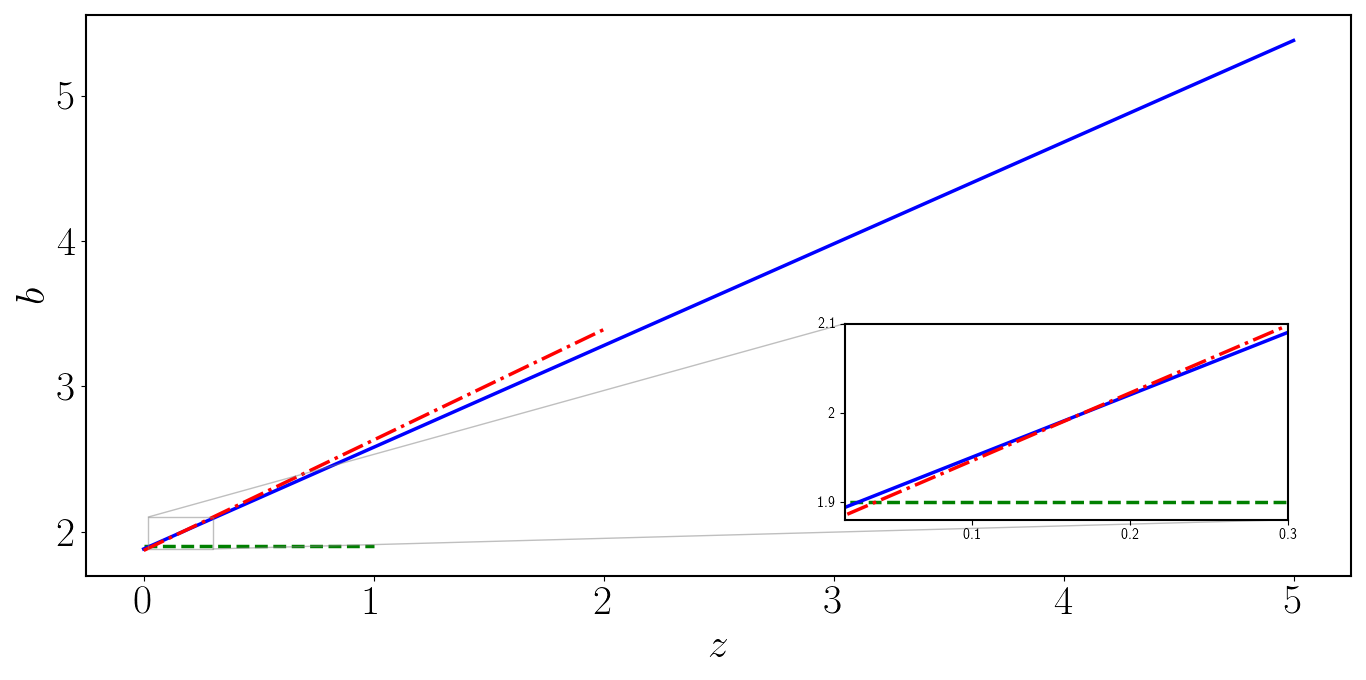}
    \caption{Fiducial bias models considered in the analysis. The color legend is the same as in fig. \ref{fig:allsources}.}
    \label{fig:allbias}
\end{figure}

To model the SN bias, we assume a constant value in the full observed redshift range (i.e., $z \in [0,1]$). This follows the conservative prescription made in \cite{Mukherjee_2020_lensing}, where SN are assumed to follow galaxy bias. The value we choose is $b_{SN} = 1.9$, which is the $b_{g}(z)$ value obtained by integrating eq. (\ref{eq:biasgal_m}) over $M_*$ and $SFR$ and averaging it between $z = 0$ and $z = 1$. In order to use the same model as in eq. (\ref{eq:bias}), the parameters $A_{SN} = 1.9$, $B_{SN} = 0$ and $P_{SN} = 0$ are defined. Fig. \ref{fig:allbias} shows the bias we obtain.

\section{Observational effects in Luminosity Distance Space}\label{sec:LDS}

When we study galaxy clustering using observational survey catalogs, we need to include corrections to account for redshift space distortions and relativistic effects, which alter the observed density field with respect to the intrinsic one (see e.g. \cite{Hamilton_1998,Yoo_2009,Challinor_2011,Breton_2018}). 
When working in Luminosity Distance Space, the situation is of course analogous. 
However, the mapping between the intrinsic density field and the observed one is different, with respect to the redshift space case. Large scale structures induce a perturbation $\epsilon$ on luminosity distance, which can be defined at leading order as follows  (cfr \cite{Hui_2006,Sasaki_1987}):
\begin{equation}\label{eq:DLpert}
    \frac{\bar{D}_L+\delta D_L}{\bar{D}_L} = 1 + \epsilon = 1+2v-\kappa \ ,
\end{equation}
where $\bar{D}_L$ is the background, unperturbed luminosity distance, $\delta D_L$ is the induced perturbation, $v$ is the radial velocity of the source along the LoS and $\kappa$ is the lensing convergence. For a given comoving distance $\chi$ and LoS $\hat{n}$, the latter is defined as:
\begin{equation}\label{eq:kappa}
    \kappa = \int_0^\chi d\chi'\frac{(\chi-\chi')\chi'}{\chi}\nabla^2_{\hat{n}}\ \psi[\chi',\hat{n} \chi'] \ , 
\end{equation}
where $\nabla^2_{\hat{n}}\ \psi[\chi',\hat{n} \chi']$ are the angular components of the covariant derivative of the Weyl potential on a sphere having radius $\chi'$.

In our previous work \cite{Libanore_2021}, we considered only the effects of LDSD coming from peculiar velocities $v$, thus neglecting $\kappa$. This was motivated by the fact that - as discussed for example in \cite{zhang2018} - peculiar velocities were expected to be the dominant source of distortion. After our previous analysis was completed, the lensing contribution was explicitly computed in \cite{Namikawa_2021}, showing that, while indeed subdominant, such contribution is not negligible on large scales.

Other effects in LDS such as gravitational potential perturbations and relativistic effects, are included neither in eq. (\ref{eq:DLpert}) nor in what follows. Such effects are still not well modelled in the literature but, despite being subdominant \cite{Hui_2006}, their analysis will be very important when considering full sky surveys and large scales, which we do not include in this work. For all these reasons, while beyond the scope of the present study, we plan on exploring these issues in a future, dedicated analysis.

\subsection{Lensing}\label{sec:lensing}

\begin{figure}[ht!]
    \centering
    \includegraphics[scale=0.4]{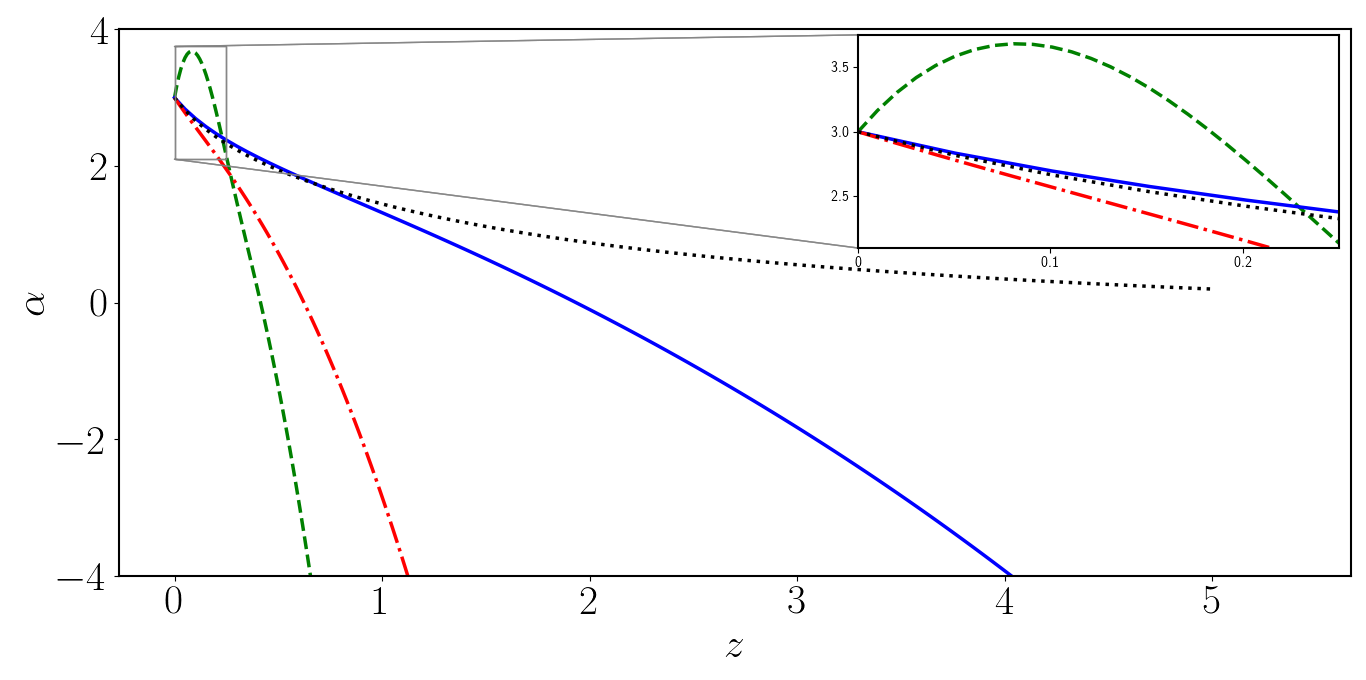} 
    \caption{$\alpha$ computed using eq. (\ref{eq:alpha}). Each curve assumes a different $n_i$ evolution in $\chi$. The distributions for SN (green dashed line), DNS (red dotted-dashed line) and DBH (blue continuous line) are computed as described in the text, while for non evolving sources (black dotted line) $d\ln n_i/d\ln\chi = 0$. BHNS are not showed for clarity; their $\alpha$ is intermediate between DBH and DNS. In the case $z = 0$, all the distributions consider $\alpha = 3$: this directly follows from the number conservation of sources between LDS and real space (see \cite{Namikawa_2021} for the complete derivation).}
    \label{fig:alpha}
\end{figure}

Our analysis of lensing contributions to LDSD starts from the results of \cite{Namikawa_2021}, which slightly differ from e.g. those reported in \cite{Hui_2006}. In particular the author of \cite{Namikawa_2021} obtains that the lensing contribution to LDSD includes both the standard term $-2\kappa$, due to the lensing distortion of the angular position, and an extra-term that depends on the LoS derivative of the lensing convergence with respect to the comoving distance. Both these terms, together with the peculiar velocity contribution, alter the observed density fluctuations $\delta_{LDS}$ with respect to the intrinsic ones $\delta$. Naming $\delta_{LDS}^{v}$ and $\delta_{LDS}^{\kappa}$ the peculiar velocity and lensing distortions on $\delta$ (i.e., propagating the terms in eq.~(\ref{eq:DLpert}) through the number conservation law as in the standard Kaiser approach \cite{kaiser_1987}), the observed density fluctuations can be described as:
\begin{equation}\label{eq:delta}
\begin{aligned}
    \delta_{LDS} &= \delta + \delta_{LDS}^v + \delta_{LDS}^{\kappa}\\
    &= \delta -2\alpha v -\frac{f_{D_L}}{aH}\frac{dv}{d\chi} + \biggl(-2+\alpha+\frac{f_{D_L}}{2aH}\frac{d}{d\chi}\biggr)\kappa \ ,
\end{aligned}
\end{equation}
where $a$ is the scale factor and $H$ represents the Hubble parameter (redshift dependencies are omitted for clarity). The parameters $f_{D_L}$ and $\alpha$ are defined as: 
\begin{align}
    f_{D_L} =&\ \frac{2\bar{D}_L}{1+z}\biggl(\frac{\partial \bar{D}_L}{\partial z}\biggr)^{-1}\ , \label{eq:fDL} \\
    \alpha =&\ \biggl[3+\frac{d\ln n_i}{d\ln\chi}-\frac{f_{D_L}}{2}+\frac{\chi f_{D_L}}{2aH}\frac{\partial(aH)}{\partial\eta}\biggr]\frac{1}{1+aH\chi}\ , \label{eq:alpha}  
\end{align}
where $\eta$ is the conformal time and $n_i$ the comoving number density of the source ($i = $ DBH, DNS, BHNS or SN). The latter is computed by dividing $N_m$ from eq. (\ref{eq:N_merger}) -- for mergers -- or $N_{SN}$ from eq. (\ref{eq:N_SN}) -- for SN -- by the comoving volume. The obtained distribution is interpolated with a skewed gaussian and analytically differentiated in order to get $d\ln n_i/d\ln\chi$. 
Eq. (\ref{eq:delta}) is the same as the one originally derived in \cite{Namikawa_2021}, where $\gamma = {f_{D_L}/2}$ is used. The term $f_{D_L}$ displays an interesting behaviour, as discussed in \cite{zhang2018,Libanore_2021}: it approaches zero at low $z$ and increases at higher redshift, crossing the value $f_{D_L} = 1$ at $z \simeq 1.7$. This means that at high redshifts ($z \gtrsim 2$), LDSD are larger than redshift space distortions, whereas at lower redshift, LDSD turn out to be very small. Since, as showed in sec. \ref{sec:SN}, the SN distribution is relevant only for $z\lesssim 1$, this implies that SN observations are almost unaffected by LDSD. This could provide interesting insights for the analysis of SN peculiar velocities, which are on their own a powerful tool in constraining Cosmology (see e.g. \cite{Davis_2011,krishnan2021hints,graziani2020peculiar}).
Fig. \ref{fig:alpha} shows the results we obtained for $\alpha$, which slightly differ from those in \cite{Namikawa_2021}, because of the different comoving number densities used in our work.

In \cite{Namikawa_2021}, it is shown that the lensing contribution to the overdensity (see eq. (\ref{eq:delta})) can be written as:
\begin{equation}
    \delta_{LDS}^{\kappa} = \int_0^{\infty}d\chi'w(\chi,\chi')\nabla_{\hat{n}}^2\ \psi[\eta(\chi'),\hat{n} \chi']\ ,
\end{equation}
where the lensing kernel $w(\chi,\chi')$ is computed using eq. (\ref{eq:kappa}), its derivative and eq. (\ref{eq:delta}): 
\begin{equation}\label{eq:lenskern}
   w(\chi,\chi') = \frac{\chi'}{\chi}\biggl[(-2+\alpha)(\chi-\chi')+\frac{f_{D_L}}{2aH}\frac{\chi'}{\chi}\biggr]\ .
\end{equation}
This expression differs from that of the ``standard'' redshift space kernel, which is  $w(\chi,\chi') = (-2 + 5s_*)(\chi-\chi')\chi'/\chi$ (cfr. \cite{Challinor_2011}). We note in particular that in this formula the $f_{DL}$ and $\alpha$ terms defined in eq. (\ref{eq:fDL}) and  (\ref{eq:alpha}) introduce the dependence on the distance from the observer, on the lensing convergence derivative along the LoS and on the source mean number density $n_i$. Moreover, the magnification bias, $s_*$, is not included. In the case of GW surveys, this was firstly computed in \cite{scelfo_2018} as: 
\begin{equation}
    s_* = -\frac{d\log\bigl[d^2N\bigl(z,\sqrt{\bigl<\rho^2\bigr>}>\rho_{lim})/dzd\Omega\bigr]}{d\sqrt{\langle\rho^2\rangle}}\Biggl|_{\rho_{lim}} \
\end{equation}
where $\rho$ is the Signal-to-Noise Ratio (SNR), $\sqrt{\langle\rho^2\rangle}$ represents its averaged value and $\rho_{lim}=8$ is the detection threshold. This term is neglected in \cite{Namikawa_2021}: we decided to keep this formulation at this level, but we plan to include a more refined model for $s_*$ in a forthcoming paper, where the full analysis of the simulated catalog and of LDSD will be performed. In the case of SN, the magnification bias is commonly neglected in literature. 

\begin{figure}[ht!]
    \centering
    \includegraphics[scale=0.4]{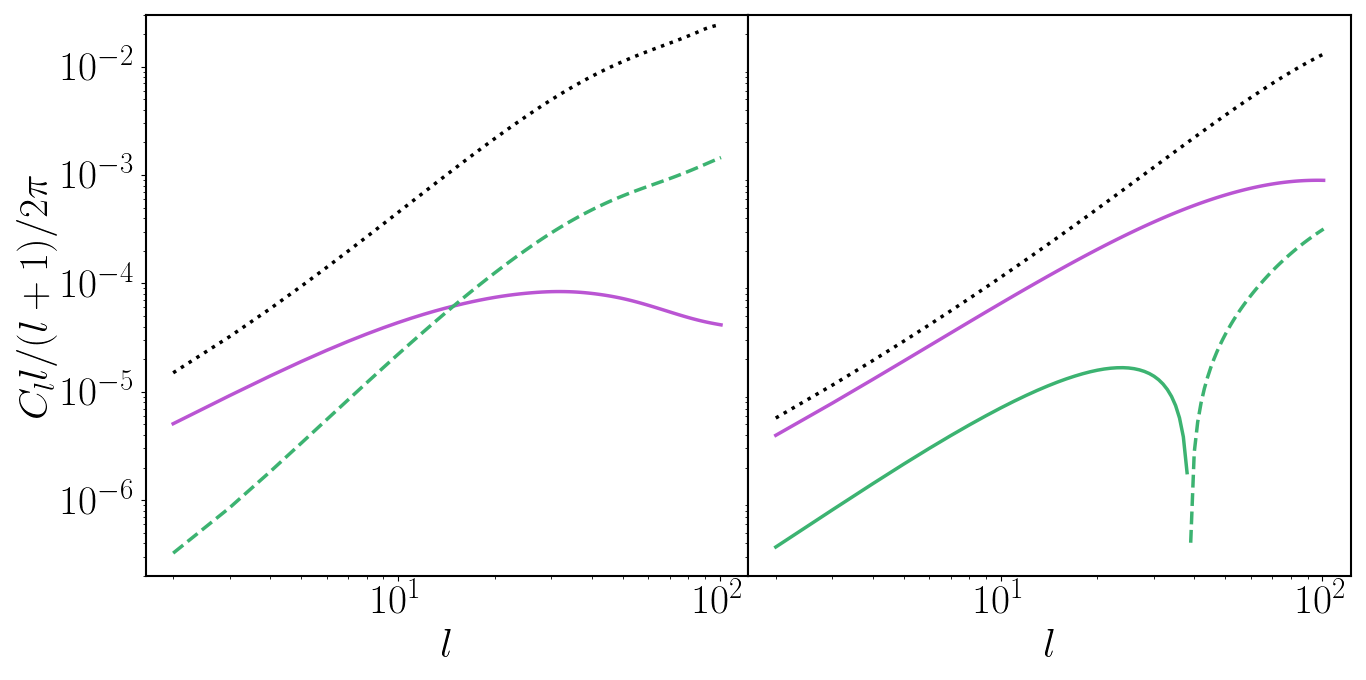}
    \caption{Overdensity (black dotted line), LDS distortions due to peculiar velocities (purple continuous line) and lensing (green continuous/dashed line) contribution to the $C_l$. Dashed lines are used for negative contributions. The left plot refers to low redshift $z \in [0.6,0.8]$, while the right one to high redshift $z \in [2.6,2.8]$. All the curves refer to the DBH distribution, without including the smoothing effect from eq. (\ref{eq:beam}).}
    \label{fig:cl_contr}
\end{figure}

\begin{figure}[ht!]
    \centering
    \includegraphics[scale=0.4]{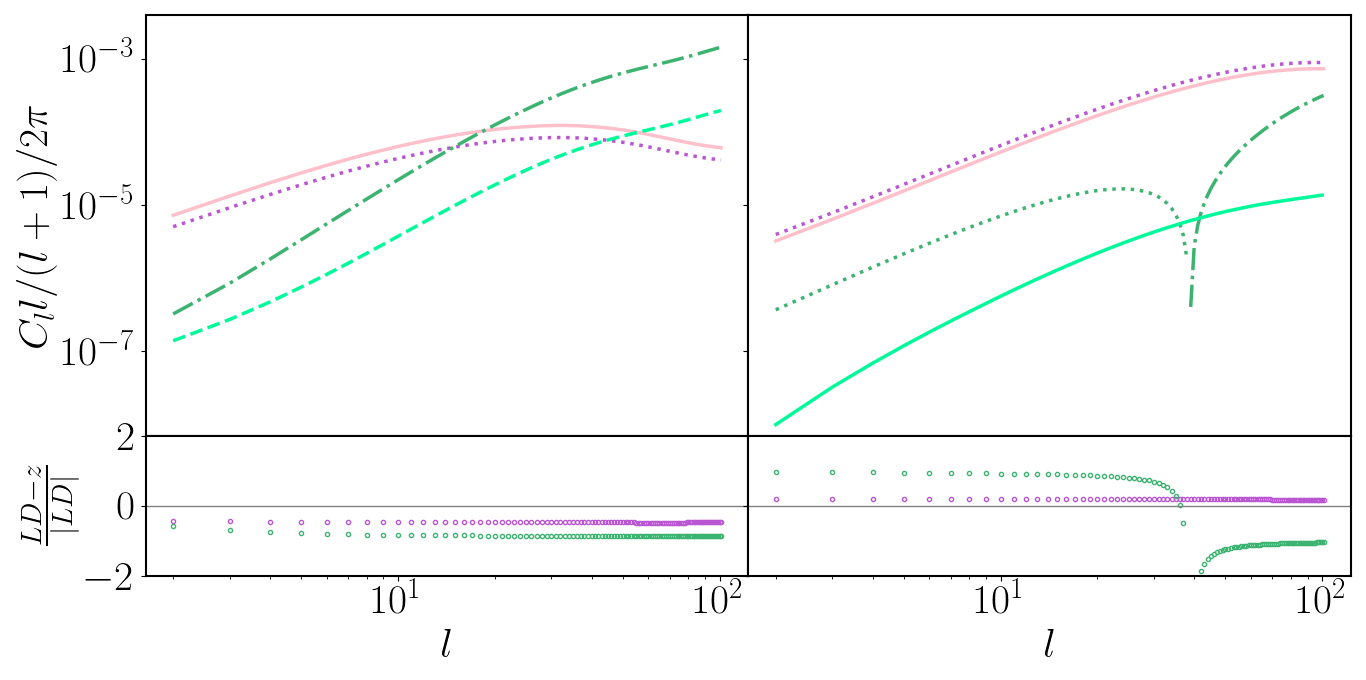}
    \caption{Comparison between peculiar velocities distortions in LDS (purple dotted line) and redshift space (pink continuous line) and between lensing in LDS (green dotted/dashed-dotted line) and redshift space (light green dotted/continuous line). Dashed or dotted-dashed lines are used for negative contributions. On the left, the DBH distribution at low redshift $z \in [0.6,0.8]$ is considered, while on the right the same at $z \in [2.6,2.8]$ is used, without including the smoothing effect described in eq. (\ref{eq:beam}). The lower plots show the residuals relative to the absolute values of quantities in LDS.}
    \label{fig:LDZ}
\end{figure}

In  \cite{Namikawa_2021}, the lensing correction is included in the computation of the two point correlation function, assuming the Limber approximation \cite{Limber_1953}. In the current analysis, we implement the new lensing kernel from eq. (\ref{eq:lenskern}) inside \texttt{CAMB}\footnote{\texttt{https://github.com/cmbant/CAMB}} \cite{Challinor_2011} -- as we did in our previous work \cite{Libanore_2021} for LDSD from peculiar velocities. Since \texttt{CAMB} assumes the Limber approximation only on very small scales (i.e. $l > 100$ in our case), our implementation accounts for full-sky effects in LDS. 
Fig. \ref{fig:cl_contr} shows the resulting contribution of peculiar velocity and lensing to LDSD, with respect to the intrinsic overdensity, in the case of low and high distances. These results are consistent with the ones from \cite{Namikawa_2021}. The shift in the lensing term when moving from redshift to LDS is bigger than the shift in the peculiar velocity case, as fig. \ref{fig:LDZ} shows; however, the lensing term still remains subdominant in LDS.

\section{Methodology}\label{sec:method}

\subsection{Multitracer analysis}

The multitracer technique was firstly developed in \cite{Seljak_2009} (see also \cite{McDonald_2009,Hamaus_2011,Abramo_2015}) to overcome the cosmic variance problem. This is relevant at large scales, where the stochastic distribution of dark matter overdensities can be measured in only a few realizations: since different tracers map the same underlying distribution, by comparing them it is possible to measure the ratio of their biases without modelling the dark matter field itself  (cfr. \cite{Witzemann_2019}). In our case, the effect of cosmic variance is shared among all the tracers and it mixes the different components of the signal \cite{abramo2013,Abramo_2015}. This improves by definition the total Signal-to-Noise Ratio and hence the forecasts on parameters as described by \cite{Seljak_2009}. Further on, our application of the multitracer technique is similar to the one used to match optical and radio datasets e.g., \cite{viljoen2021multiwavelength}, where the enhancement of the constraints on bias sensitive parameters are a straightforward consequence of the multitracer Fisher matrix. This is the case of LDSD and any other effect that is manifested through differences between the clustering of distinct species of tracers. Although Cosmology and bias are not fully degenerate, they are correlated \cite{zheng2007,zhao2021completed}, and hence improving the constraints of the bias parameters increases the cosmological power of clustering measurements, indirectly improving the constraints through parameter correlations.

Our forecast analysis expands and improves over our previous work \cite{Libanore_2021} both by including the lensing perturbation described in sec. \ref{sec:lensing} and by applying the multitracer technique. The multitracer analysis in our case brings together the GW survey performed by ET/ET$\times$3 and VRO observations of SN. As a further case, we also check the results for the combination of different merger events e.g., DBH and DNS. Despite being performed with the same detector (i.e., ET/ET$\times$3), GW from different mergers can be considered as observed by independent surveys since their signal can be easily distinguished. For this reason, we refer as well to combinations of different kinds of mergers with the name of "multitracer analysis".

In order to test the stability of our results, we compute the Fisher matrix in two different, yet equivalent, ways. First of all, we consider the Fisher matrix for the overdensity field, which is Gaussian distributed with zero average:
\begin{equation}\label{eq:fisherstandard}
    F_{\alpha\beta} = \sum_l \frac{2l+1}{2}f_{sky}\text{Tr}\bigl[\partial_{\alpha}\mathbf{C}_l\ \Gamma_l^{-1}\partial_{\beta}\mathbf{C}_l\ \Gamma_l^{-1}\bigr] \ , 
\end{equation}
where $f_{sky}$ is the observed sky fraction and $\partial_{\alpha,\beta}\mathbf{C}_l$ is the matrix of the derivatives with respect to the parameters of interest (see sec. \ref{sec:der} for details) of the auto- and cross- power spectra computed in the different distance bins. These are defined as:
\begin{equation}
    C_l^{ij} = 4\pi \int d\ln k\ P^{pr}(k)\ \Delta^{W}_{N,l}(D_L^i,k)\Delta^{W}_{N,l}(D_L^i,k) \ , 
\end{equation}
where $k$ is the scale in Fourier space, $P^{pr}(k) = A_s(k/k_0)^{n_s-1}$ the primordial power spectrum ($k_0$ is the pivot scale) and $\Delta^{W}_{N,l}(D_L^{i,j},k)$ are the observational window functions, defined in each $D_L$ bin centered in $D_L^i$ as:
\begin{equation}
    \Delta^{W}_{N,l}(D_L^{i,j},k) = \int dD_L\  p(D_L)W(D_L^i,D_L)\Delta_{l}(D_L,k) \ , 
\end{equation}
being $p(D_L) = \bigl[d^2N/dzd\Omega\bigr] / \int dD_L\ p(D_L)W(D_L^i,D_L)$ the normalised observed number of sources, $W(D_L^i,D_L)$ the Gaussian window function centered in $D_L^i$, and $\Delta_{l}(D_L,k)$ the theoretical transfer function. In the LDS framework (see sec. \ref{sec:LDS}), considering the source $s_a$, this (cfr. \cite{Challinor_2011,fonseca19,Namikawa_2021}) is given\footnote{The expression in eq. (\ref{eq:transfer}) is approximated, since we are neglecting relativistic effects (see sec. \ref{sec:LDS}).} in terms of the Bessel functions $j_l(k\chi)$ and their derivatives by: 
\begin{equation}\label{eq:transfer}
\begin{aligned}
    \Delta_{l}(D_L,k) &= b_{s_a}\delta_k^{DM}j_l(k\chi)+f_{D_L}\frac{kv_k}{aH}j_l^{''}(k\chi)\ + \\
    & + \frac{l(l+1)}{2} \int_0^{\chi} d\tilde{\chi}\ \frac{\chi'}{\tilde{\chi}}\biggl[(-2+\alpha)(\tilde{\chi}-\chi')+\frac{f_{D_L}/2}{aH}\frac{\chi'}{\tilde{\chi}}\biggr]\bigl[\phi_k(\tilde{\chi})+\psi_k(\tilde{\chi})\bigr]j_l(k\tilde{\chi}) \ .
\end{aligned}
\end{equation}
In the previous equation, $b_{s_a}$ is the source bias, $\chi$ the comoving distance, $\delta_k^{DM}$ the DM overdensity, $v_k,\ \phi_k,\ \psi_k$ respectively the peculiar velocity, the gravitational potential and the Bardeen potential in Fourier space. All the other quantities are defined in sec. \ref{sec:LDS}. The difference between eq. (\ref{eq:transfer}) and its standard expression in redshift space consists in the presence of the $f_{D_L}$ factor and in the different lensing kernel, which was defined in eq. (\ref{eq:lenskern}).

In eq. (\ref{eq:fisherstandard}), $\Gamma_l = \mathbf{C}_l+\mathbf{N}_l$ is the field covariance matrix, including noise contributions. We  assume only shot noise contributions, uncorrelated between different bins and tracers i.e., $N_{ij}^{s_a,s_b}=\delta^{K}_{s_a,s_b}\delta^{K}_{ij}[N_{s_a}(z_i)]^{-1}$ (being $s_{a,b} = $ SN or DNS, BHNS, DBH) . We construct $\mathbf{C}_l$ (and analogously $\Gamma_l$) as a block matrix combining all the tracers and distance bins.\footnote{For example, in the case of $2$ tracers and $2$ bins $\mathbf{C}_l$ is defined as:
\begin{equation}
    \mathbf{C}_l = \left(
\begin{array}{cc|cc}
C_{00}^{s_1,s_1} & C_{00}^{s_1,s_2} & C_{01}^{s_1,s_1} & C_{01}^{s_1,s_2}  \\
C_{00}^{s_2,s_1} & C_{00}^{s_2,s_2} & C_{01}^{s_2,s_1} & C_{01}^{s_2,s_2} \\
\midrule
C_{10}^{s_1,s_1} & C_{10}^{s_1,s_2} & C_{11}^{s_1,s_1} & C_{11}^{s_1,s_2}\\
C_{10}^{s_2,s_1} & C_{10}^{s_2,s_2} & C_{11}^{s_2,s_1} & C_{11}^{s_2,s_2}\\
\end{array}
\right)_l
\end{equation}
}

We then take an "estimator perspective", i.e., we consider as our actual observables the estimates $\hat{C}_l$ of the various auto- and cross- power spectra, rather than the overdensity field. Power spectrum estimates are then characterized by a Wishart sampling distribution. Under a Gaussian approximation, one can write the corresponding Fisher matrix as \cite{Carron_2013,euclid_2021f}:
\begin{equation}\label{eq:fisher}
    F_{\alpha\beta} = \sum_l F_{\alpha\beta, l} = \sum_l (2l+1)f_{sky}\ \bigl[(\partial_{\alpha}\mathsf{C}_l)^{T}\ \Gamma_l^{'-1}\ \partial_{\beta}\mathsf{C}_l \bigl]\ , 
\end{equation}
$\mathsf{C}_l = \bigl[\hat{C}_{00}^{s_1,s_2}\ \hat{C}_{00}^{s_1,s_2}\ ...\bigr]_l$ being the vector containing all the auto- and cross- power spectra and $\Gamma_l^{'}$ being the power spectrum covariance matrix, defined as:
\begin{equation}
\begin{aligned}
    \Gamma_l' &= \text{cov}\bigl[\hat{C}_{ij}^{s_a,s_b}\hat{C}_{km}^{s_c,s_d}\bigr]_l \\
   &= \bigl[(C_{ij}^{s_a,s_b} + N_{ij}^{s_a,s_b})(C_{km}^{s_c,s_d} + N_{km}^{s_c,s_d})+
    (C_{ik}^{s_a,s_c} + N_{ik}^{s_a,s_c})(C_{jm}^{s_c,s_d} + N_{jm}^{s_c,s_d})\bigr]_l \ .
\end{aligned}
\end{equation}

\noindent
The number of spectra involved in the computation scales as $n_a^2n_{i}^2$, where  $n_a$, $n_i$ are the number of sources and the number of bins used in the analysis, respectively.\footnote{Considering for example $1$ tracer in $2$ bins, the terms $F_{\alpha\beta,l}$ in eq. (\ref{eq:fisher}) would be rewritten as:
\begin{equation}
\begin{aligned}
    F_{\alpha\beta,l} &= \sum_l (2l+1)f_{sky}
    \left[\begin{array}{cccc}
\hat{C}_{00}^{s_1,s_1} & \hat{C}_{01}^{s_1,s_1} & \hat{C}_{10}^{s_1,s_1} & \hat{C}_{11}^{s_1,s_1} 
\end{array}
\right]
\times \\
\times & \left[
\begin{array}{cccc}
\text{cov}[\hat{C}_{00}^{s_1,s_1}\hat{C}_{00}^{s_1,s_1}] & \text{cov}[\hat{C}_{00}^{s_1,s_1}\hat{C}_{01}^{s_1,s_1}] & \text{cov}[\hat{C}_{00}^{s_1,s_1}\hat{C}_{10}^{s_1,s_1}] & \text{cov}[\hat{C}_{00}^{s_1,s_1}\hat{C}_{11}^{s_1,s_1}]  \\
\text{cov}[\hat{C}_{01}^{s_1,s_1}\hat{C}_{00}^{s_1,s_1}] & \text{cov}[\hat{C}_{01}^{s_1,s_1}\hat{C}_{01}^{s_1,s_1}] & \text{cov}[\hat{C}_{01}^{s_1,s_1}\hat{C}_{10}^{s_1,s_1}] & 
\text{cov}[\hat{C}_{01}^{s_1,s_1}\hat{C}_{11}^{s_1,s_1}] \\
\text{cov}[\hat{C}_{10}^{s_1,s_1}\hat{C}_{00}^{s_1,s_1}] & \text{cov}[\hat{C}_{10}^{s_1,s_1}\hat{C}_{01}^{s_1,s_1}] & \text{cov}[\hat{C}_{10}^{s_1,s_1}\hat{C}_{10}^{s_1,s_1}] & \text{cov}[\hat{C}_{10}^{s_1,s_1}\hat{C}_{11}^{s_1,s_1}]  \\
\text{cov}[\hat{C}_{11}^{s_1,s_1}\hat{C}_{00}^{s_1,s_1}] & \text{cov}[\hat{C}_{11}^{s_1,s_1}\hat{C}_{01}^{s_1,s_1}] & \text{cov}[\hat{C}_{11}^{s_1,s_1}\hat{C}_{10}^{s_1,s_1}] & 
\text{cov}[\hat{C}_{11}^{s_1,s_1}\hat{C}_{11}^{s_1,s_1}] \\
\end{array}
\right]
\times \left[
\begin{array}{c}
\hat{C}_{00}^{s_1,s_1} \\ \hat{C}_{01}^{s_1,s_1} \\ \hat{C}_{10}^{s_1,s_1} \\ \hat{C}_{11}^{s_1,s_1} 
\end{array}
\right] .
\end{aligned}
\end{equation}
}

We checked that changing $F$ from eq. (\ref{eq:fisherstandard}) to eq. (\ref{eq:fisher}) produces fully consistent results. All the results described in sec. \ref{sec:results_single} and \ref{sec:results_multi} are obtained using $F$ from eq. (\ref{eq:fisherstandard}).

\subsection{Survey specifications and analysis framework}
Auto- and cross- angular power spectra used to get results in sec. \ref{sec:results} are tomographically computed for the different tracer combinations reported in tab. \ref{tab:tracer_combo}. For each survey we assume the specifications described in tab. \ref{tab:survey}: for SN, we rely on the VRO survey, while for mergers we check both the single ET scenario and the ET$\times$3 scenario that we introduced in \cite{Libanore_2021}. The only difference between the two resides in the sky localization uncertainty, which in the ET$\times$3 case allows us to probe smaller scales; it is interesting to explore whether similar sky localization improvements could possibly be obtained using statistical techniques or cross-correlations with galaxies.  

\begin{table}[ht!]
    \centering
    \small
    \begin{tabular}{|c|l|l|}
    \toprule
    Analysis type & Survey & Tracer \\
    \midrule
\multirow{2}{*}{Single tracer}  & VRO & SN \\
& ET & DNS, BHNS, DBH \\
& ET$\times$3 & DNS, BHNS, DBH \\
    \midrule
\multirow{2}{*}{Multitracer}    & ET$\times$3 & DNS + DBH \\
    & ET $+$ VRO & DBH + SN \\
    & ET$\times$3 $+$ VRO & DBH + SN \\
    \bottomrule
    \end{tabular}
    \caption{Single and multitracer combinations considered in the analysis. As sec. \ref{sec:results_single} describes, VRO is the Vera Rubin Observatory, ET stands for Einstein Telescope, and ET$\times$3 for a network made by 3 ET detectors. For each tracer the survey specifications are reported in tab. \ref{tab:survey}. BHNS are not included in the multitracer analysis, as sec. \ref{sec:results_multi} explains.}
    \label{tab:tracer_combo}
\end{table}

\begin{table}[ht!]
\small
\centering
\begin{tabular}{|c|c|c|c|c|c|c|c|}
\toprule
Survey & Source & Area $[$deg$^2]$ & $\Delta D_L/D_L$ & $\Delta\Omega$ $[$deg$^2]$ & $z_{max}$ & $T^{OBS}$ &  $N^{TOT}$ \\
\midrule
VRO & SN & $18000$ & $0.15$ & $\sim 0$ & $1$ & 5yr & $10^{4.86}$\\
\midrule 
\multirow{3}{*}{ET } & DNS & \multirow{3}{*}{Full sky} & $0.3$ & \multirow{3}{*}{$100$} & $2$ & \multirow{3}{*}{3yr}&  $10^{4.14}$\\
& BHNS & & $0.3$ & & $3$ & & $10^{4.37}$ \\
& DBH & & $0.1$ & & $5$ & & $10^{4.79}$ \\
\midrule 
\multirow{3}{*}{ET$\times$3} & DNS & \multirow{3}{*}{Full sky} & $0.3$ & $10$ & $2$ & \multirow{3}{*}{3yr}&  $10^{4.14}$\\
& BHNS & & $0.3$ & $10$ & $3$ & & $10^{4.37}$ \\
& DBH & & $0.1$ & $3$ & $5$ & & $10^{4.79}$ \\
\bottomrule
\end{tabular}
\caption{Survey specifications (specs): observed sky area, luminosity distance and sky localization uncertainties ($\Delta D_L/D_L$, $\Delta\Omega$, respectively), detector horizon $z_{max}$. VRO specs are in accordance with \cite{Howlett2017}; the sky localization is extremely precise for SN, therefore we assume $\Delta\Omega \sim 0$. ET specs are compatible with \cite{Zhao_2018} (where a full description depending on frequency and distance for DNS and BHNS is provided) and \cite{vitale_2017} (where a full analysis for DBH can be found), while for ET$\times$3 we assumed the same specs as in \cite{Libanore_2021}; the only difference between ET and ET$\times$3 relies in $\Delta\Omega$. $N^{TOT}$ is computed as described in sec. \ref{sec:tracer} assuming $T^{OBS}$.}
\label{tab:survey}
\end{table}

\begin{figure}[ht!]
    \centering
    \includegraphics[scale=0.37]{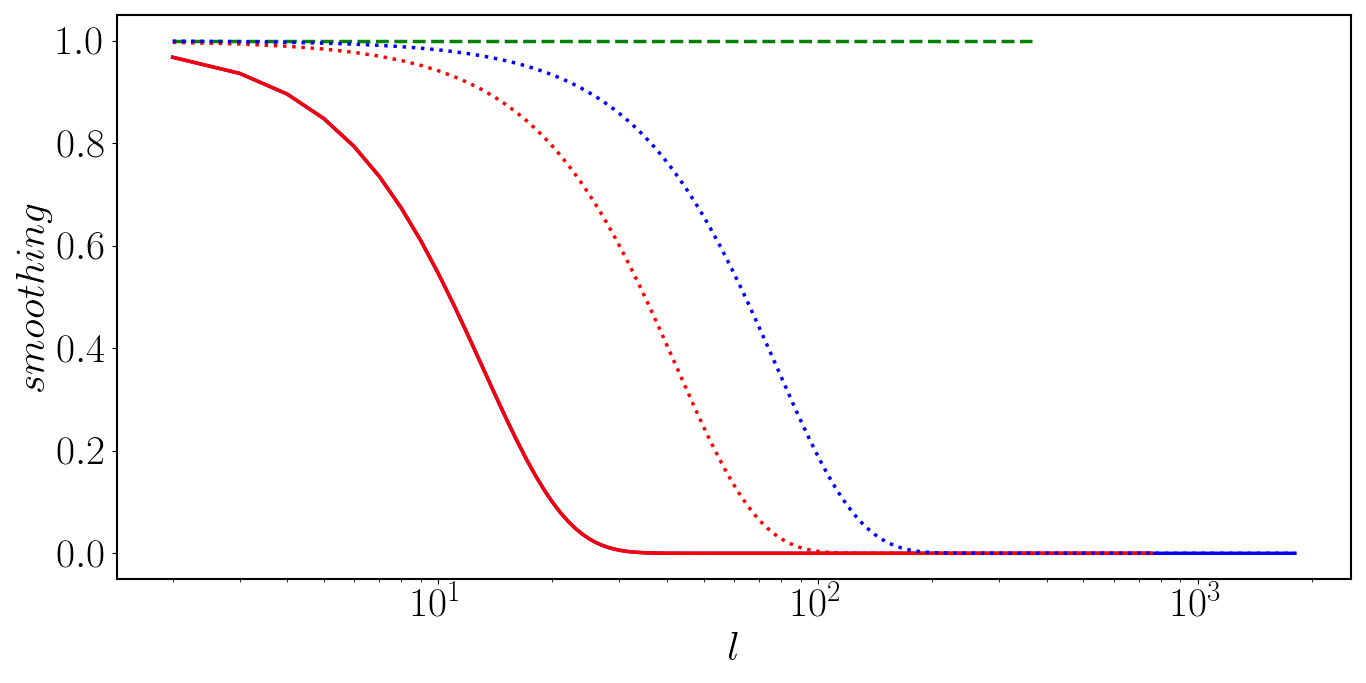}
    \caption{Beam window defined through eq. (\ref{eq:beam}) for SN (green dashed line) and sources observed through ET (red continuous line for DNS, overlapping for DBH and BHNS). The ET$\times$3 scenario is described by the red dotted line (DNS) and blue dotted (DBH) line. Each line is computed considering $l_{max}$  as the one associated through eq. (\ref{eq:lmax}) to $z_{max}$ reached for the source (see tab. \ref{tab:survey}). }
    \label{fig:beam}
\end{figure}

At each redshift, we consider only modes in the linear regime. More specifically we choose a cut-off:
\begin{equation}\label{eq:lmax}
    l_{max}(z_i,z_j) = k_{nl}^0\min\bigl[(1+z_{i,j})^{2/(2+n_s)}\chi(z_{i,j})\bigr] \ , 
\end{equation}
where $k_{nl}^0= 0.1\ h\text{Mpc}^{-1}$ is the non-linear cut-off scale at $z= 0$. 
We finally account for the sky localization uncertainty $\Delta\Omega$ by considering for each couple of tracers $s_a,\ s_b$, a Gaussian beam profile with $\sigma_{s_{a,b}} = \sqrt{\Delta\Omega_{s_{a,b}}/8\ln{2}}$, with $\Delta\Omega_{s_{a,b}}$ from tab. \ref{tab:survey} and converted to sterradians:
\begin{equation}\label{eq:beam}
    B^{s_{a}s_{b}}_l = \exp{\biggl[{-\frac{l(l+1)}{2}\sigma_{s_a}^2}\biggr]}\exp{\biggl[{-\frac{l(l+1)}{2}\sigma_{s_b}^2}\biggr]} \ .
\end{equation}
In the case of SN, since the sky localization is extremely precise compared to GW observations, we assume $\Delta\Omega_{SN} \sim 0$. Fig. \ref{fig:beam} shows the smoothing eq. (\ref{eq:beam}) provides on the different surveys. 

\subsection{Parameters of interest}\label{sec:der}
In our Fisher analysis, we consider the following set of parameters (i.e., $\alpha,\ \beta$ in eq. (\ref{eq:fisher})): 
\begin{itemize}
    \item [\tiny $\bullet$] Cosmological parameters: $[H_0$, $\Omega_ch^2$, $w_0$, $w_a$, $\Omega_bh^2$, $n_s$, $A_s]$. Fiducial values are the ones from {\it Planck 2018} \cite{Planck2018}.\footnote{As for the dark energy Equation of State, we assumed the Chevallier-Polarski-Linder (CPL) parametrization $w(a) = w_0 + w_a(1-a)$ to be consistent with the standard \texttt{CAMB} configuration. Other authors (see e.g \cite{DDE}) prefer to use different models, as they claim the CPL expansion to be poorly sensitive at low $z$. Alternative parametrizations could be used to constrain the dynamical dark energy paradigm: implementing them in our code would straightforwardly provide forecasts through an analysis analogous to the CPL one.}
    \item [\tiny $\bullet$] A pair of bias related parameters $[A_i,P_i]$ for each source $i =$ SN, DNS, BHNS, DBH (see eq. (\ref{eq:bias})). $A_i$ and $P_i$ represent respectively the amplitude and the slope of the bias. The parameter $B_i$ in eq. (\ref{eq:bias}), which describes the bias local value, is considioned to the value extracted from the simulations. The reason for this choice is that, using as fiducial values the ones described in sec. \ref{sec:tracer} and including both $A_i$, $B_i$ in our analysis would make the constraints on the two fully degenerate. As described in sec. \ref{sec:prior}, in the case of DBH we also check the effect of including a Gaussian prior on the slope parameter $P_{DBH}$ (owing to the fact that results from simulations strongly indicate a linear dependence of the $b_i(z)$ on redshift).
\end{itemize}

\noindent
Derivatives are computed numerically for cosmological parameters, analytically for bias.

\section{Results}\label{sec:results}

As tab. \ref{tab:tracer_combo} shows, we analyse different surveys, both considering one tracer (sec. \ref{sec:results_single}) and combining many of them to apply the multitracer technique (sec. \ref{sec:results_multi}). The specifications assumed for each kind of survey are described in tab. \ref{tab:survey}. Marginalized errors are obtained for cosmological and bias parameters using the Fisher matrix formalism described in sec. \ref{sec:method}; uniform priors are assumed for all the parameters, except in sec. (\ref{sec:prior}) where the effects of a prior on the bias slope are analysed. We checked that including {\it Planck 2018} \cite{Planck2018} priors on cosmological parameters does not produce a significant improvement in the final constraints in the multitracer analysis with ET$\times$3, whereas in a single tracer GW analysis, the {\it Planck} prior would simply dominate over the signal.

\subsection{Single tracer analysis}\label{sec:results_single}

Tab. \ref{tab:cosmo_single} and tab. \ref{tab:bias_single} show the forecasted marginalized $1-\sigma$ errors for cosmological and bias parameters, respectively. 

BHNS mergers do not produce any significant improvement over DNS constraints, despite the slightly larger number of these events (cfr. tab. \ref{tab:survey}) at slightly higher redshifts ($2<z<3$). For this reason, we do not explicitly report BHNS forecasts in tab. \ref{tab:cosmo_single} and \ref{tab:bias_single} and we do not include them in the multitracer analysis in the next section.
\begin{table}[ht!]
    \centering
    \begin{tabular}{|c|c|c|c|c|}
    \toprule 
    & $H_0$ & $\Omega_ch^2$ & $w_0$ & $w_a$ \\
    \midrule
    SN & $7.864$ & $1.168 \cdot10^{-2}$ & $1.070 \cdot10^{-1}$ & $9.828 \cdot10^{-1}$ \\
    DNS ET$\times$3 & $76.67$ & $1.470 \cdot10^{-2}$ & $1.798$ & $12.23$ \\
    DBH ET$\times$x3 &  $16.28$ & $2.446 \cdot10^{-2}$ & $2.552 \cdot10^{-1}$ & $1.622$ \\
    \bottomrule
    \end{tabular}
    \begin{tabular}{|c|c|c|c|}
    & $\Omega_bh^2$ & $n_s$ & $A_s$\\
    \midrule
    SN & $1.468 \cdot10^{-3}$ & $8.924 \cdot10^{-2}$ & $3.315 \cdot10^{-10}$ \\
    DNS ET$\times$3 & $2.651 \cdot 10^{-2}$ & $3.613 \cdot10^{-1}$ & $ 2.947 \cdot10^{-9}$ \\
    DBH ET$\times$3 & $5.134\cdot 10^{-3}$ & $1.783 \cdot10^{-1}$ & $5.976 \cdot10^{-10}$ \\ 
    \bottomrule
    \end{tabular}
    \caption{Marginalized $1-\sigma$ errors for cosmological parameters, single tracer analysis.}
    \label{tab:cosmo_single}
\end{table}

\begin{figure}[ht!]
    \centering
    \includegraphics[scale=0.4]{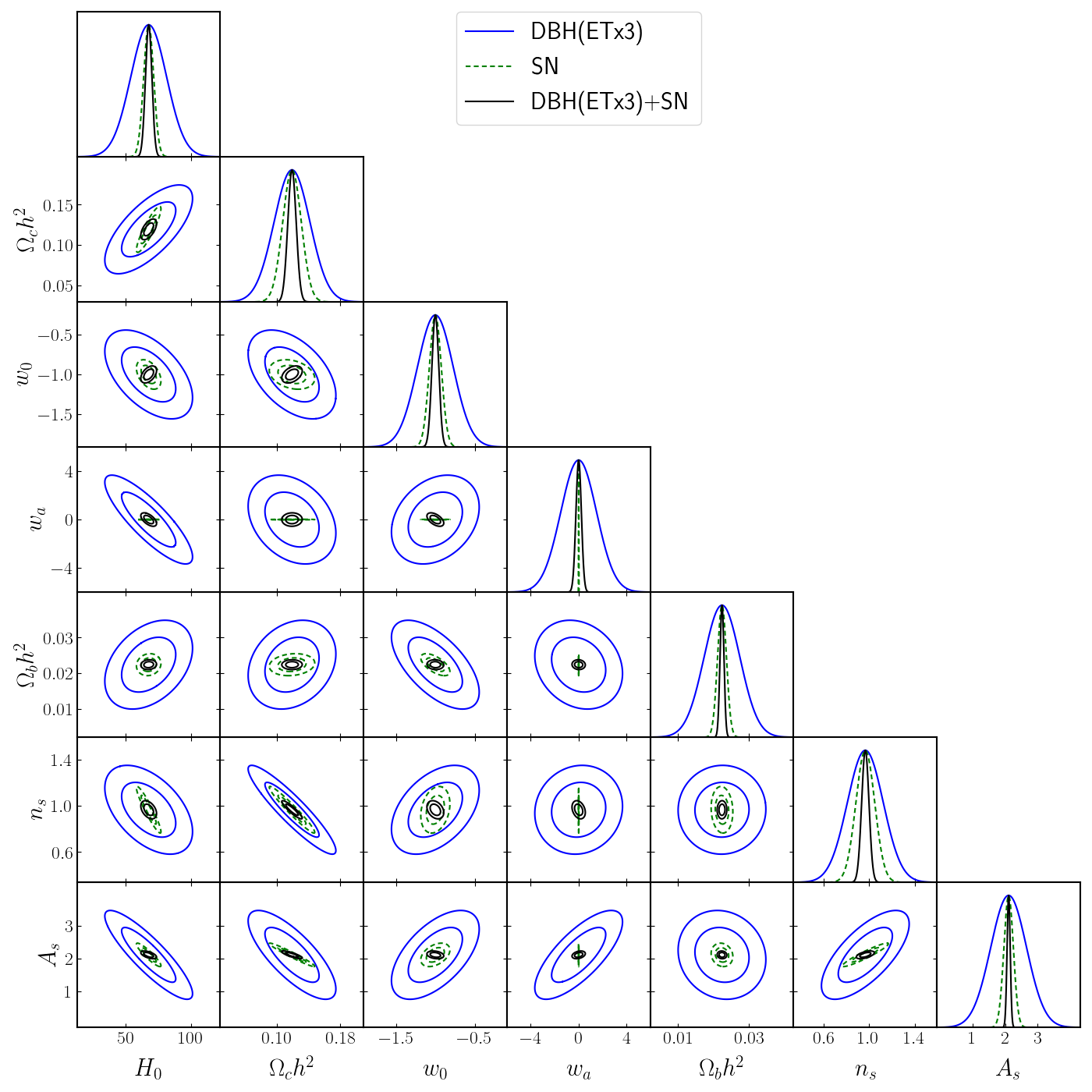}
    \caption{Confidence $1-$ and $2-\sigma$ ellipses for cosmological parameters computed for DBH (ET$\times$3, blue continuous line), SN (green dashed line) and DBH(ET$\times$3)$+$SN (black continuous line).}
    \label{fig:newellipse}
\end{figure}

As for cosmological parameters, the low-precision sky localization assumed for ET prevents us from having good constraints on cosmological parameters for both DNS and DBH. As we also discussed in \cite{Libanore_2021}, considering ET$\times$3 leads to significant improvements, even though the overall constraining power remains relatively small, compared to other cosmological probes.
Forecasts for SN are much better instead, since the VRO allows us to probe smaller, but still linear scales, compared to GW. Fig. \ref{fig:newellipse} compares confidence ellipses for DBH, SN and for the multitracer analysis (see sec. \ref{sec:results_multi}) in constraining the cosmological parameters.

\begin{figure}[ht!]
    \centering
    \includegraphics[scale=0.4]{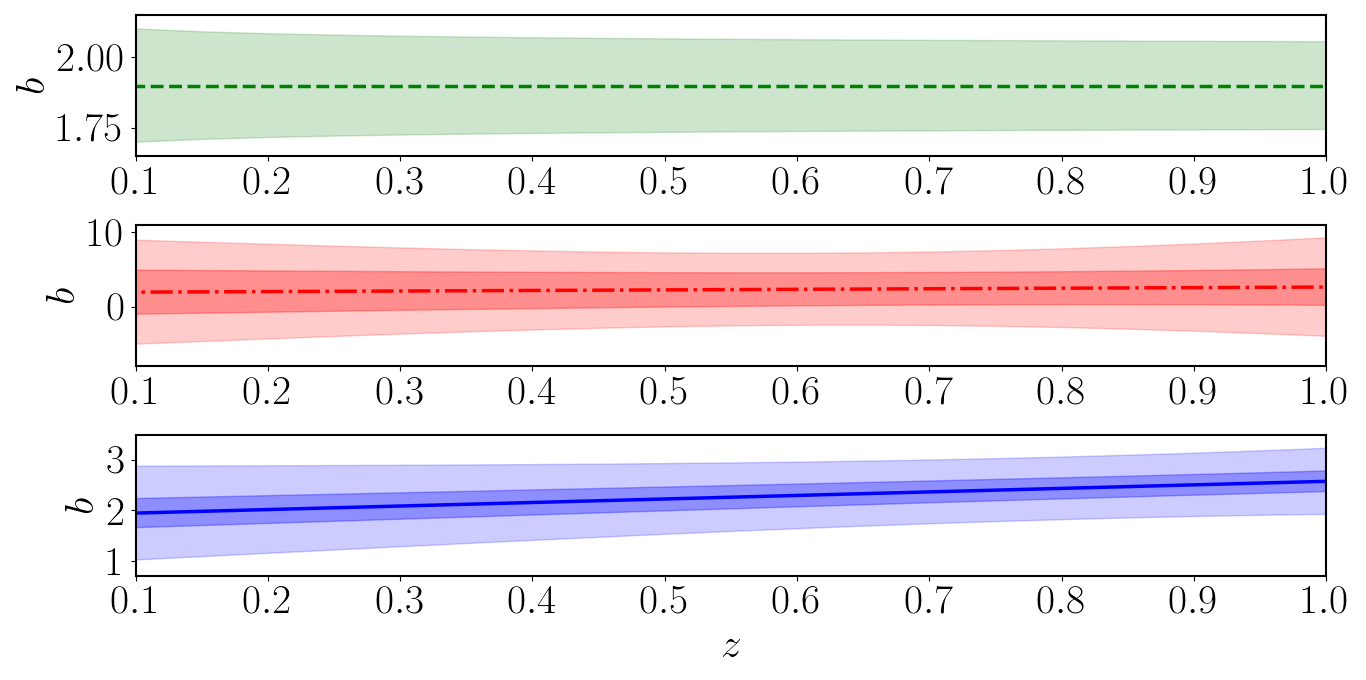}
    \caption{Bias marginalized $1-\sigma$ errors at low $z$ for SN (green dashed line, upper panel), DNS (red dotted-dashed line, central panel) and DBH (blue continuous line, lower panel) single tracer analysis. The bold line represent the fiducial models described in sec. \ref{sec:tracer}, while the shaded areas are included between $b_i\pm\sigma_{b_i}$, with $\sigma_{b_i}$ from eq. (\ref{eq:berr}). For DNS and DBH both the ET (lighter) and ET$\times$3 (darker) constraints are showed. }
    \label{fig:bias_forecast}
\end{figure}

We now turn our attention to bias parameters ($A_i$, $P_i$), and show in fig. \ref{fig:bias_forecast} the uncertainties which they produce on the estimates of the bias coefficients, $b_i(z)$. Since the bias for each source is computed as $b_i(z) = A_i(z+B_i)^{P_i}$ (where $B_i$ is held fixed at 2.68 for DBH, 2.46 for DNS and 0 for SN, as described in sec. \ref{sec:tracer}), $\sigma_{b_i}$ is computed by error propagation:
\begin{equation}\label{eq:berr}
    \sigma_{b_i}^2 = 
    \begin{bmatrix} \dfrac{\partial b_i}{\partial A_i} & \dfrac{\partial b_i}{\partial P_i} \end{bmatrix} 
    \begin{bmatrix} \sigma_{A_i}^2 & \text{cov}(A_i,P_i) \\ \text{cov}(A_i,P_i) & \sigma_{P_i}^2 \end{bmatrix} \begin{bmatrix} \dfrac{\partial b_i}{\partial A_i} \\ \dfrac{\partial b_i}{\partial P_i} \end{bmatrix} \ .
\end{equation}
The values of $\sigma_{A_i}$ and $\sigma_{P_i}$ are described in tab. \ref{tab:bias_single}. The single tracer analysis provides very good constraints for SN bias, since its errors relative to the fiducial is $\sigma_{b_{SN}(z)}/b_{SN} \lesssim 10\%$ over all the redshifts considered. 
As for GW events instead, both ET and ET$\times$3 forecasts are a bit worse with respect to our previous work \cite{Libanore_2021}: this depends on the degeneracy between the amplitude and slope parameters; this is further studied in sec. \ref{sec:prior}, where ways to lift such degeneracy are also discussed.

\begin{table}[ht!]
    \centering
    \begin{tabular}{|c|c|c||c|c||c|c|}
    \toprule 
    & $A_{DBH}$ & $P_{DBH}$ & $A_{DNS}$ & $P_{DNS}$ & $A_{SN}$ & $P_{SN}$\\
    \midrule
    SN  & $-$ & $-$ & $-$ & $-$ & $0.156$ & $0.020$ \\
    DNS & $-$ & $-$ & $12.80$ & $14.65$ & $-$ & $-$\\
    DNS ET$\times$3 & $-$ & $-$ & $4.572$ & $5.014$ & $-$ & $-$\\
    DBH & $1.446$ & $1.607$ & $-$ & $-$ & $-$ & $-$\\ 
    DBH ET$\times$3 & $0.382$ & $0.404$ & $-$ & $-$ & $-$ & $-$\\ 
    \bottomrule
    \end{tabular}
    \caption{Marginalized $1-\sigma$ errors for bias parameters, single tracer analysis.}
    \label{tab:bias_single}
\end{table}

\subsection{Multitracer analysis}\label{sec:results_multi}

In our multitracer analysis we consider different combinations of tracers and detectors, to test their impact on the final forecasts. The use of two or more tracers makes it possible to cover a larger redshift range (with the density distributions of different sources peaking at different $z$), add cross-power spectrum information, break degeneracies and lower the effect of cosmic variance, through the different bias dependence on redshift \cite{Seljak_2009}. 

Tab. \ref{tab:cosmo_multi} and \ref{tab:bias_multi} show the forecasted $1-\sigma$ marginalized errors obtained respectively for cosmological and bias parameters. We decided not to show the analysis for DNS$+$BHNS$+$ DBH$+$SN since the DNS and BHNS distributions do not introduce noticeable improvements. 
\begin{table}[ht!]
    \centering
    \begin{tabular}{|c|c|c|c|c|}
    \toprule 
    & $H_0$ & $\Omega_ch^2$ & $w_0$ & $w_a$ \\
    \midrule
   DNS $+$ DBH (ET$\times$3) & $7.472$ & $1.987\cdot10^{-2}$ & $1.938\cdot10^{-1} $ & $1.035$ \\
   DBH (ET) $+$ SN & $6.386$ & $8.378 \cdot10^{-3}$ & $9.135\cdot10^{-2}$ & $1.158 \cdot10^{-1}$ \\
   DBH (ET$\times$3) $+$ SN & $3.012$ & $6.983 \cdot10^{-3}$ & $6.701\cdot10^{-2}$ & $2.532 \cdot10^{-1}$ \\
    \bottomrule
    \end{tabular}
    \begin{tabular}{|c|c|c|c|}
    & $\Omega_bh^2$ & $n_s$ & $A_s$\\
    \midrule
   DNS $+$ DBH (ET$\times$3) & $3.772 \cdot10^{-3}$ & $1.603\cdot10^{-1}$ & $3.405\cdot10^{-10}$  \\
   DBH (ET) $+$ SN & $2.108 \cdot10^{-3}$ & $6.716 \cdot10^{-2}$ & $1.023 \cdot10^{-10}$ \\
   DBH (ET$\times$3) $+$ SN & $5.675 \cdot10^{-4}$ & $5.554 \cdot10^{-2}$ & $5.842 \cdot10^{-11}$ \\
    \bottomrule
    \end{tabular}
    \caption{Marginalized $1-\sigma$ errors for cosmological parameters, multitracer analysis. }
    \label{tab:cosmo_multi}
\end{table}

On one hand, the DNS$+$DBH(ET$\times$3) scenario only slightly improves the forecasts with respect to sec. \ref{sec:results_single}: this happens because DNS and DBH bias models are close one to the other (cfr fig. \ref{fig:allbias}) and therefore their combination can not disentangle the power spectra dependence on the different parameters.
On the other hand, when including SN, forecasts largely improve. The SN contribution is dominant with respect to the DBH one when considering a single ET detector, making a joint analysis in this case not very meaningful: forecasts for DBH(ET)$+$SN are in fact very close to those obtained in the single-tracer, SN case. If we account for ET$\times$3, instead, the parameter sensitivity achievable with either DBH or SN alone is more comparable. Therefore, the multitracer analysis allows us in this case to significantly improve the final constraints, over the single tracer scenario.
As a benchmark, we can compare results from the multitracer DBH(ET$\times$3)$+$SN case with those obtainable for galaxy clustering in a future {\it Euclid}-like survey \cite{euclid_2020}. Our multitracer forecasts are comparable with the {\it Euclid}-like ones for most of the parameters, the main exceptions being $H_0$ -- for which GW$+$SN have quite poor constraining power -- and dark energy parameters $w_0$ and $w_a$ -- for which we obtain somewhat tighter constraints in the GW $+$ SN analysis, thanks to the small scales reached by SN combined with the large volumes probed by DBH. 

\begin{figure}[ht!]
    \centering
    \includegraphics[scale=0.4]{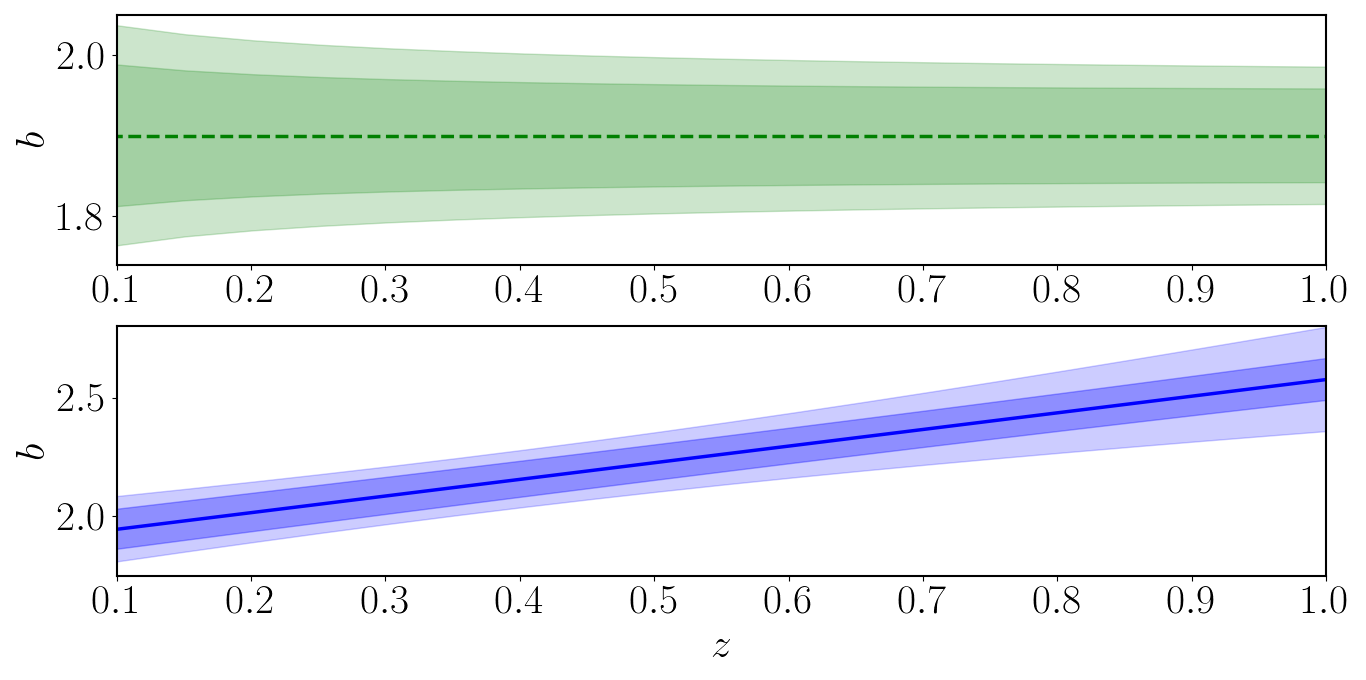}
    \caption{Bias error forecasts for SN and DBH in the DBH(ET)$+$SN and DBH(ET$\times$3)$+$SN cases; the legend is the same of fig. \ref{fig:bias_forecast}.}
    \label{fig:bias_forecast_multi}
\end{figure}

If we now turn our attention to bias parameter constraints, we see again significant improvements coming from the multitracer approach, considering both ET and ET$\times$3. Tab. \ref{tab:bias_multi} shows that the constraining power on $A_i$ and $P_i$ (hence on $b_i(z)$) improves considerably with respect to the single tracer scenario, especially when we consider DBH$+$SN analysis (the same would happen for DNS$+$SN). In particular, in the DBH(ET)$+$SN scenario, $\sigma_{b_{DBH}(z)}/b_{DBH} \lesssim 35\%$ over all the distribution, while for DBH(ET$\times$3)$+$SN $\sigma_{b_{DBH}(z)}/b_{DBH} \lesssim 15\%$. The improvement in the estimate of the DBH bias over all the redshift range is evident if we compare fig. \ref{fig:bias_forecast} and fig. \ref{fig:bias_forecast_multi}.
Also in the case of bias parameters, since DBH and DNS bias fiducial models are very similar, the DNS+DBH combination provides forecasts for $A_{DBH},\ P_{DBH},\ A_{DNS},\ P_{DNS}$ that are similar to those obtained with the single tracer analysis in tab. \ref{tab:bias_single}. 

\begin{table}[ht!]
    \centering
    \begin{tabular}{|c|c|c||c|c||c|c|}
    \toprule 
    & $A_{DBH}$ & $P_{DBH}$ & $A_{DNS}$ & $P_{DNS}$ & $A_{SN}$ & $P_{SN}$\\
    \midrule
  DNS $+$ DBH (ET$\times$3) & $0.303$ & $0.322$ & $0.379$ & $0.456$ & $-$ & $-$\\
  DBH (ET) $+$ SN & $0.308$ & $0.384$ & $-$ & $-$ & $0.085$ & $0.018$\\
  DBH (ET$\times$3) $+$ SN & $0.126$ & $0.145 $ & $-$ & $-$ & $0.058$ & $ 0.013$\\
    \bottomrule
    \end{tabular}
    \caption{Marginalized $1-\sigma$ errors for bias parameters, multitracer analysis.}
    \label{tab:bias_multi}
\end{table}

\subsection{Effects of a prior on bias parameters}\label{sec:prior}
The analysis of merger bias is particularly interesting to understand clustering properties of binary mergers and disentangle different formation scenarios. Since the ET catalogs will be dominated by DBH observations, we are particularly interested in understanding the detectability of their bias. As described in sec. \ref{sec:results_single} and \ref{sec:results_multi}, the bias error $\sigma_{b_{DBH}}$ depends on both the uncertainties on the amplitude $A_{DBH}$ and on the slope $P_{DBH}$ parameters. The two, however, are degenerate one with respect to the other and the degeneracy worsens the capability of this kind of analysis in detecting the bias. 

This can be checked by comparing single tracer bias forecasts described in sec. \ref{sec:results_single} with results from our previous paper \cite{Libanore_2021}. In our previous work, the use of different bias parameters in each $D_L$ bin allowed us to detect bias at low redshift. Using the $[A_{DBH},P_{DBH}]$ parameterization both in the single ET and the ET$\times$3 makes $b_{DBH}(z)$ however always undetectable in the single-tracer part of the current analysis. This issue can actually be effectively solved by using the multitracer technique, as sec. \ref{sec:results_multi} shows.
Moreover, we note that in sec. \ref{sec:results_single} and \ref{sec:results_multi} we assume uninformative priors on both $A_{DBH}$ and $P_{DBH}$. 
However, in our analysis in sec. \ref{sec:tracer} the slope parameter turned out to be well described by a linear trend: we obtained this result through an HOD analysis of hydrodynamical simulations \cite{Artale2018,Artale2019}, but also other works in the literature, despite following different approaches, found the same behaviour, at least up to $z \sim 3.5$ (see e.g. \cite{Scelfo_2020,Mukherjee_2021}). For this reason, it seems reasonable to increase the level of "reliability" of the fiducial value $P_{DBH} = 1$ by associating a Gaussian prior to it. 

\begin{figure}[ht!]
    \centering
    \includegraphics[scale=0.37]{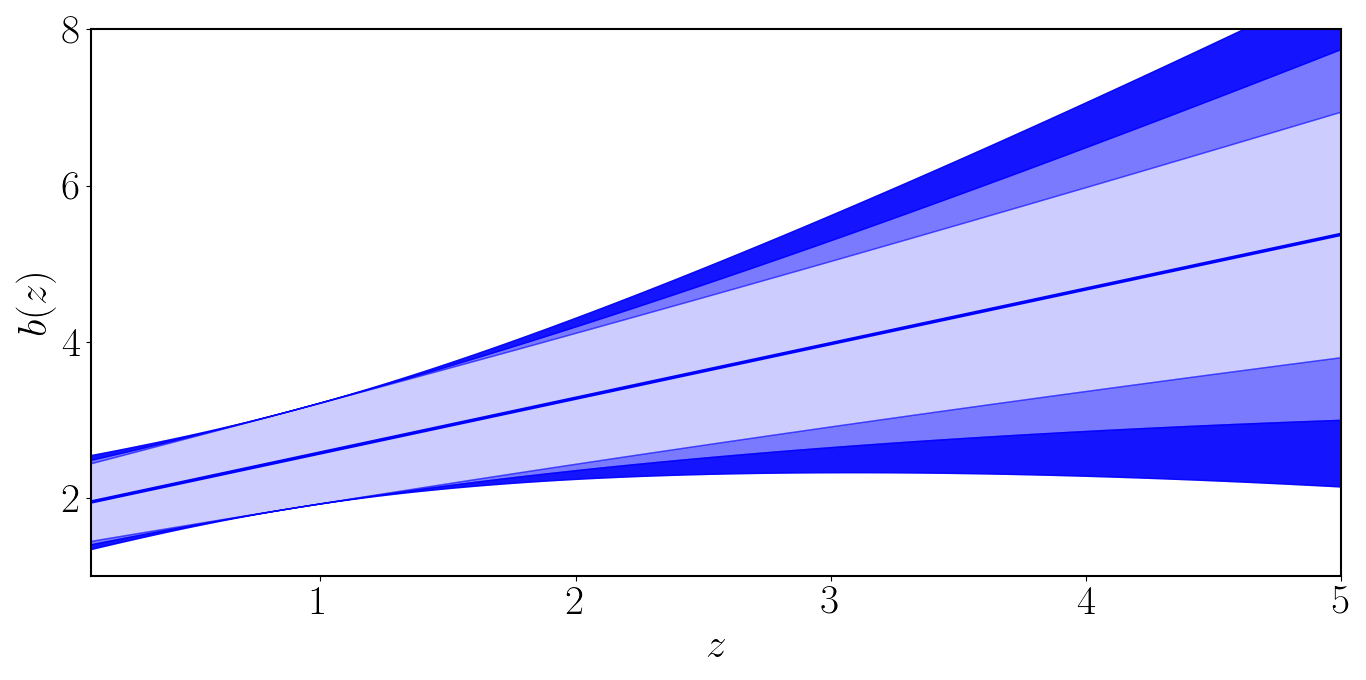}\\
    \includegraphics[scale=0.37]{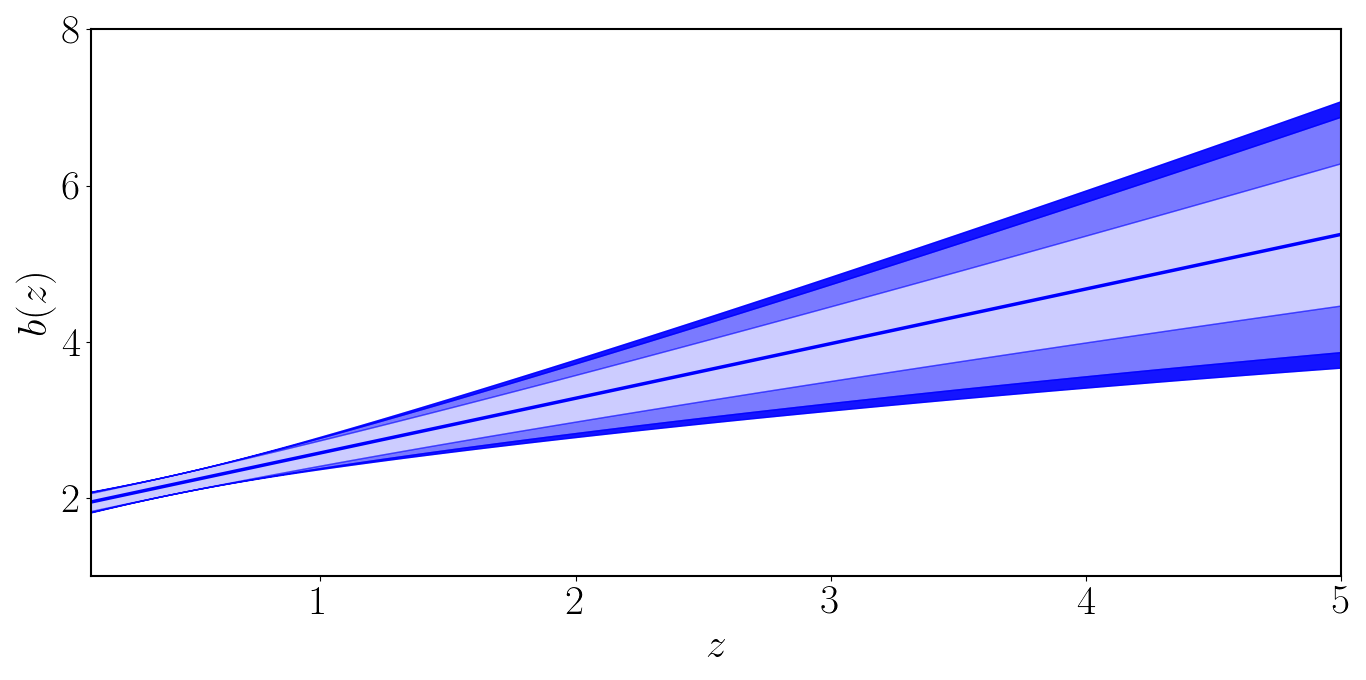}
    \caption{DBH bias error obtained in the DBH, single ET case (upper plot) and in the DBH(ET)$+$SN case (lower plot) assuming the different priors on $P_{DBH}$ described in tab. \ref{tab:prior}. The darker area is related to the larger prior ($80\%$), the intermediate color to $50\%$ and the lighter area is the one related with the tighter prior (i.e., $20\%$).}
    \label{fig:prior}
\end{figure}

Tab. \ref{tab:prior} shows the effect of including different $P_{DBH}$ priors in the single and full multitracer cases. All the measurements refer to the single ET or ET$\times$3 scenarios. As expected, the more information the survey already provides, the less impact the prior inclusion has in the analysis.
Fig. \ref{fig:prior} shows how the error propagation through eq. (\ref{eq:berr}) changes depending on the $P_{DBH}$ prior in the DBH and DBH+SN cases, both considering a single ET detector. Assuming $\sigma_{P_{DBH}}^{prior}/b_{DBH} = 50\%$ in the prior allows us to achieve $\sigma_{b_{DBH}}/b_{DBH} \lesssim 50\%$ in the single-tracer DBH case, while in the DBH(ET)$+$SN case we obtain $\sigma_{b_{DBH}}/b_{DBH} \lesssim 25\%$, over the entire redshift range. These constraints obviously further improve considering ET$\times$3, as tab. \ref{tab:prior} shows.

\begin{table}[ht!]
    \centering
    \begin{tabular}{|c|c|c|c||c|c|}
    \toprule
    \multirow{2}{*}{$\sigma_{P_{DBH}}^{prior}/P_{DBH}$} &  \multirow{2}{*}{tracer} & \multicolumn{2}{c||}{ET} & \multicolumn{2}{c|}{ET$\times$3} \\
    & & $A_{DBH}$ & $P_{DBH}$& $A_{DBH}$ & $P_{DBH}$ \\
    \midrule
    \multirow{2}{*}{$100\%$} & DBH & $1.446$ & $1.607$ & $0.382$ & $0.404$ \\
    & DBH $+$ SN & $0.308$ & $0.384$  & $0.126$ & $0.145$\\
    \midrule
    \multirow{2}{*}{$80\%$} & DBH & $0.663$ & $0.716$ & $0.341$ & $0.361$ \\
    & DBH $+$ SN & $0.278$& $0.346$ & $0.124$ & $0.142$ \\
    \midrule
    \multirow{2}{*}{$50\%$} & DBH & $0.461$ & $0.477$ & $0.299$ & $0.314$ \\
    & DBH $+$ SN & $0.245$& $0.305$ & $0.121$ & $0.139$ \\
    \midrule
    \multirow{2}{*}{$20\%$} & DBH & $0.249$ & $0.198$ & $0.176$ & $0.179$ \\
    & DBH $+$ SN & $0.146$& $0.177$ & $0.103$ & $0.117$ \\
    \bottomrule
    \end{tabular}
    \caption{Effects of different $P_{DBH}$ Gaussian priors on the marginalized errors of bias parameters. The first line sums up results from tab. \ref{tab:bias_single} and \ref{tab:bias_multi}, where uninformative, uniform prior is considered.}
    \label{tab:prior}
\end{table}

\section{Conclusions}

In this work, we studied the power spectrum of different cosmic tracers in Luminosity Distance Space. Our goal was that of studying the constraining power of these observables on cosmological and bias parameters, by means of a Fisher matrix approach. We considered future GW and SN observations, such as those which will be made by Einstein Telescope (either in a single detector configuration or in a network with $3$ detectors) and the Vera Rubin Observatory.

In a previous analysis \cite{Libanore_2021} we performed a similar forecast, but we considered only GW mergers, in a single tracer approach. Consistently with our previous analysis, we showed that the lack of sky localization in the single ET scenario prevents us from observing scales small enough to provide good constraints on cosmological parameters, while ET$\times$3 can do much better. Using a multitracer analysis in which Supernovae are combined with GW observations from ET$\times$3 leads to further significant improvements.\footnote{In the single ET scenario, SN have instead a completely dominant constraining power over GW, making the multitracer approach not so useful.}
The main source of such improvements can be traced to the breaking of significant degeneracies between different bias coefficients and cosmological parameters.
This happens because the information encoded in the SN and DBH datasets are complementary: SN probe lower redshifts, while DBH probe large volumes and very high redshifts. In the majority of cases, DNS mergers give a less important contribution to this type of analysis, due to the lower resolution in $D_L$ of their dataset and to the fact that DNS do not probe the high redshift region, where only DBH can be found. Moreover, the degeneracy between DNS and DBH bias makes the combination DNS$+$DBH only slightly more predictive, compared to the single-tracer analysis for either of them.

Our main results for cosmological parameters are shown in tab. \ref{tab:cosmo_multi}. If, as a benchmark test, we compare the DBH(ET$\times$3)$+$SN multitracer results to the forecasts for a future {\it Euclid}-like survey for galaxy clustering, reported in \cite{euclid_2020}, we find similar constraining power in for most parameters, with the exception of $H_0$ -- which is constrained poorly by DBH(ET$\times$3)$+$SN -- and of dark energy parameters, $w_0$ and $w_a$ -- which are instead constrained better by DBH(ET$\times$3)$+$SN.
These results might seem a little surprising at first sight, because the expected number of galaxies in forthcoming surveys, such as {\it Euclid}, is expected to be significantly larger than the number of GW mergers or SN, even considering third generation GW experiments (as we are doing here with ET). However, this is compensated by the much larger volumes covered in GW merger surveys, especially when considering DBH events, combined with the increase in the number of tracers when we combine DBH and SN. Independently of the relative constraining power of galaxy and GW merger clustering, the most interesting aspect is the complementarity between different datasets. Moreover, the analysis of merger and SN clustering in LDS presents specific peculiarities, compared to the redshift space analysis of galaxy clustering, which are further discussed below and in sec. \ref{sec:LDS}. 

In a cosmological parameter analysis, bias parameters are just treated as nuisance parameters and marginalized out. GW merger bias coefficients, however, are interesting per se, since they can provide insights into the nature of the physical processes underlying the mergers themselves. For this reason, a study of the detectability of $b_{i}(z)$, with $i =$ DBH, DNS, BHNS, after marginalizing out the cosmological parameters, is also interesting \cite{Scelfo_2020}. Our fiducial model for merger bias is based on the outputs of hydrodynamical N-body simulations \cite{Artale2018,Artale2019}, processed via an HOD approach (see sec. \ref{sec:GW} and \cite{Libanore_2021}). The main results of the bias forecast analysis have been summarized in fig. \ref{fig:prior}: in the case of GW sources, when considering the single tracer analysis the degeneracy between the amplitude $A_i$ and slope $P_i$ parameters makes the bias detection challenging. We showed however that in the case of DBH both the multitracer technique (sec. \ref{sec:method}) and the inclusion of a Gaussian prior on the slope (sec. \ref{sec:prior}) make the bias well detectable even at high redshift. This improves the results of our previous work \cite{Libanore_2021}, where bias turned out to be detectable only at low redshift.

Besides the inclusion of SN and the combination of different tracers, another significant new aspect of our analysis is the inclusion of lensing contributions in the treatment of distortions in Luminosity Distance Space. 
In our previous analysis, \cite{Libanore_2021} we neglected lensing terms in LDSD  because, according to \cite{zhang2018}, they were expected to be subdominant with respect to peculiar velocity contributions. 
In the meantime, an explicit calculation of lensing terms in the LDSD expression at first order was performed in \cite{Namikawa_2021}, where a numerical evaluation of the resulting GW merger power spectra was obtained in Limber approximation and it was shown that the lensing contribution was indeed smaller, but not negligible with respect to the peculiar velocity one, especially on large scales. In this work, we modified the publicly available \texttt{CAMB} code \cite{Challinor_2011} to implement the LDSD formula of \cite{Namikawa_2021}, including lensing terms and without resorting to any flat-sky approximation. The effects of these additional terms were then propagated to the final parameter forecasts, showing that their impact is indeed very small for this specific analysis. A full treatment of all GR effects on very large scales, in luminosity distance space, is still missing in the literature and will be pursued in a forthcoming publication \cite{bertacca_prep}. When these will be available, it will be possible to study GW surveys properly even on the largest scales: by doing so, it will be interesting to test the forecasting power of ET in measuring the non-Gaussianity parameter $f_{NL}$, via the merger power spectrum at low-$k$. This would be in fact another suitable target for this kind of surveys: since only large scales are required, on one hand, full sky observations are needed, while, on the other, it is not necessary to have a good sky localization. We are currently developing studies on the constraining power of ET on $f_{NL}$ and we will provide our results in a future, dedicated work.

\acknowledgments
The authors thank D. Bertacca, A. Ricciardone and A. Raccanelli for useful discussions. We thank A. Ghosh for the internal LIGO/Virgo Collaboration review and the anonymous referee for the comments that helped us clarifying our analysis.
\noindent ML was supported by the project "Combining Cosmic Microwave Background and Large Scale Structure data: an Integrated Approach for Addressing Fundamental Questions in Cosmology", funded by the MIUR Progetti di Ricerca di Rilevante Interesse Nazionale (PRIN) Bando 2017 - grant 2017YJYZAH. ML, NB and SM acknowledge partial financial support by ASI Grant No. 2016-24-H.0. MM and YB acknowledge financial support from the European Research Council for the ERC Consolidator grant DEMOBLACK, under contract no. 770017. 
MCA and MM acknowledge financial support from the Austrian National Science Foundation through FWF stand-alone grant P31154-N27. 
DK acknowledge financial support from  the South African Radio Astronomy Observatory (SARAO) and the National Research Foundation (Grant No. 75415).

\bibliography{biblio}

\begin{thebibliography}{64}
\providecommand{\natexlab}[1]{#1}
\providecommand{\url}[1]{\texttt{#1}}
\expandafter\ifx\csname urlstyle\endcsname\relax
  \providecommand{\doi}[1]{doi: #1}\else
  \providecommand{\doi}{doi: \begingroup \urlstyle{rm}\Url}\fi

\bibitem[Abbott~et al.(2021{\natexlab{a}})]{abbottGWTC2.1}
R.~Abbott~et al.
\newblock {GWTC-2.1: Deep Extended Catalog of Compact Binary Coalescences
  Observed by LIGO and Virgo During the First Half of the Third Observing Run},
  2021{\natexlab{a}}.

\bibitem[Abbott~et al.(2021{\natexlab{b}})]{abbottO3a}
R.~Abbott~et al.
\newblock {GWTC-2: Compact Binary Coalescences Observed by LIGO and Virgo
  during the First Half of the Third Observing Run}.
\newblock \emph{Physical Review X}, 11\penalty0 (2), 2021{\natexlab{b}}.
\newblock \doi{10.1103/PhysRevX.11.021053}.

\bibitem[Abbott~et al.(2021{\natexlab{c}})]{abbott_2021}
R.~Abbott~et al.
\newblock Observation of gravitational waves from two neutron
  star{\textendash}black hole coalescences.
\newblock \emph{The Astrophysical Journal Letters}, 915\penalty0 (1),
  2021{\natexlab{c}}.
\newblock \doi{10.3847/2041-8213/ac082e}.

\bibitem[Abramo and Leonard(2013)]{abramo2013}
L.~R. Abramo and K.~E. Leonard.
\newblock Why multitracer surveys beat cosmic variance.
\newblock \emph{Monthly Notices of the Royal Astronomical Society},
  432\penalty0 (1):\penalty0 318–326, Apr 2013.
\newblock ISSN 1365-2966.
\newblock \doi{10.1093/mnras/stt465}.

\bibitem[Abramo~et al.(2016)]{Abramo_2015}
L.~Abramo~et al.
\newblock Fourier analysis of multitracer cosmological surveys.
\newblock \emph{Mon. Not. Roy. Astron. Soc.}, 455\penalty0 (4), 2016.

\bibitem[Artale~et al.(2018)]{Artale2018}
M.C. Artale~et al.
\newblock {The impact of assembly bias on the halo occupation in hydrodynamical
  simulations}.
\newblock \emph{Monthly Notices of the Royal Astronomical Society},
  480\penalty0 (3):\penalty0 3978--3992, 2018.

\bibitem[Artale~et al.(2019)]{Artale2019}
M.C. Artale~et al.
\newblock {Mass and star formation rate of the host galaxies of compact binary
  mergers across cosmic time}.
\newblock \emph{Monthly Notices of the Royal Astronomical Society},
  491\penalty0 (3):\penalty0 3419--3434, 2019.

\bibitem[Belgacem~et al.(2018{\natexlab{a}})]{Belgacem2_2018}
E.~Belgacem~et al.
\newblock Modified gravitational-wave propagation and standard sirens.
\newblock \emph{Physical Review D}, 98\penalty0 (2), 2018{\natexlab{a}}.

\bibitem[Belgacem~et al.(2018{\natexlab{b}})]{Belgacem_2018}
E.~Belgacem~et al.
\newblock Gravitational-wave luminosity distance in modified gravity theories.
\newblock \emph{Physical Review D}, 97\penalty0 (10), 2018{\natexlab{b}}.

\bibitem[Bertacca~et al.(in prep.)]{bertacca_prep}
D.~Bertacca~et al.
\newblock {}, in prep.

\bibitem[Blanchard~et al.(2020)]{euclid_2020}
A.~Blanchard~et al.
\newblock Euclid preparation.
\newblock \emph{Astronomy $\&$ Astrophysics}, 642, 2020.
\newblock \doi{10.1051/0004-6361/202038071}.

\bibitem[Breton~et al.(2018)]{Breton_2018}
M.~A. Breton~et al.
\newblock Imprints of relativistic effects on the asymmetry of the halo
  cross-correlation function: from linear to non-linear scales.
\newblock \emph{Monthly Notices of the Royal Astronomical Society},
  483\penalty0 (2), 2018.
\newblock \doi{10.1093/mnras/sty3206}.

\bibitem[Calore~et al.(2020)]{CAlore_2020}
F.~Calore~et al.
\newblock Cross-correlating galaxy catalogs and gravitational waves: A
  tomographic approach.
\newblock \emph{Phys. Rev. Research}, 2, 2020.
\newblock \doi{10.1103/PhysRevResearch.2.023314}.

\bibitem[Carron(2013)]{Carron_2013}
J.~Carron.
\newblock On the assumption of gaussianity for cosmological two-point
  statistics and parameter dependent covariance matrices.
\newblock \emph{Astronomy $\&$ Astrophysics}, 551, 2013.
\newblock ISSN 1432-0746.
\newblock \doi{10.1051/0004-6361/201220538}.

\bibitem[Cañas-Herrera et~al.(2021)Cañas-Herrera, Contigiani, and
  Vardanyan]{Ca_as_Herrera_2021}
G.~Cañas-Herrera, O.~Contigiani, and V.~Vardanyan.
\newblock Learning how to surf: Reconstructing the propagation and origin of
  gravitational waves with gaussian processes.
\newblock \emph{The Astrophysical Journal}, 918\penalty0 (1), 2021.
\newblock ISSN 1538-4357.
\newblock \doi{10.3847/1538-4357/ac09e3}.

\bibitem[Challinor and Lewis(2011)]{Challinor_2011}
A.~Challinor and A.~Lewis.
\newblock Linear power spectrum of observed source number counts.
\newblock \emph{Physical Review D}, 84\penalty0 (4), 2011.
\newblock ISSN 1550-2368.

\bibitem[Collaboration(2020{\natexlab{a}})]{Abbott_2020}
LIGO~Scientific Collaboration.
\newblock Prospects for observing and localizing gravitational-wave transients
  with advanced ligo, advanced virgo and kagra.
\newblock \emph{Living Reviews in Relativity}, 23\penalty0 (1),
  2020{\natexlab{a}}.

\bibitem[Collaboration and Collaboration(2016)]{LIGO2016}
LIGO~Scientific Collaboration and Virgo Collaboration.
\newblock Observation of gravitational waves from a binary black hole merger.
\newblock \emph{Phys. Rev. Lett.}, 116:\penalty0 061102, 2016.

\bibitem[Collaboration(2009)]{LSST20019}
LSST~Science Collaboration.
\newblock Lsst science book, version 2.0, 2009.

\bibitem[Collaboration(2020{\natexlab{b}})]{Planck2018}
Planck Collaboration.
\newblock Planck 2018 results - vi. cosmological parameters.
\newblock \emph{Astronomy and Astrophysics}, 641, 2020{\natexlab{b}}.
\newblock ISSN 1432-0746.

\bibitem[Cremonese et~al.(2021)Cremonese, Ezquiaga, and
  Salzano]{Cremonese_2021}
P.~Cremonese, J.~M. Ezquiaga, and V.~Salzano.
\newblock Breaking the mass-sheet degeneracy with gravitational wave
  interference in lensed events.
\newblock \emph{Physical Review D}, 104\penalty0 (2), 2021.
\newblock \doi{10.1103/physrevd.104.023503}.

\bibitem[Cullan~et al.(2017)]{Howlett2017}
H.~Cullan~et al.
\newblock Measuring the growth rate of structure with type {IA} supernovae from
  {LSST}.
\newblock \emph{The Astrophysical Journal}, 847\penalty0 (2), sep 2017.

\bibitem[Dalang~et al.(2020)]{Dalang_2020}
C.~Dalang~et al.
\newblock Horndeski gravity and standard sirens.
\newblock \emph{Physical Review D}, 102\penalty0 (4), 2020.

\bibitem[Davis~et al.(2011)]{Davis_2011}
T.M. Davis~et al.
\newblock The effect of peculiar velocities on supernova cosmology.
\newblock \emph{The Astrophysical Journal}, 741\penalty0 (1):\penalty0 67, oct
  2011.
\newblock \doi{10.1088/0004-637x/741/1/67}.

\bibitem[Diaz and Mukherjee(2021)]{muk_2021_bias}
C.~C. Diaz and S.~Mukherjee.
\newblock {Mapping the cosmic expansion history from LIGO-Virgo-KAGRA in
  synergy with DESI and SPHEREx}.
\newblock \emph{arXiv:2107.12787}, 7 2021.

\bibitem[Fonseca et~al.(2019)Fonseca, Viljoen, and Maartens]{fonseca19}
José Fonseca, Jan-Albert Viljoen, and Roy Maartens.
\newblock Constraints on the growth rate using the observed galaxy power
  spectrum.
\newblock \emph{Journal of Cosmology and Astroparticle Physics}, 2019\penalty0
  (12):\penalty0 028–028, Dec 2019.
\newblock ISSN 1475-7516.
\newblock \doi{10.1088/1475-7516/2019/12/028}.

\bibitem[Garcia~et al.(2020)]{Garcia2020}
K.~Garcia~et al.
\newblock On the amount of peculiar velocity field information in supernovae
  from lsst and beyond.
\newblock \emph{Physics of the Dark Universe}, 29:\penalty0 100519, 2020.

\bibitem[Garoffolo~et al.(2020)]{Garoffolo_2020}
A.~Garoffolo~et al.
\newblock Gravitational waves and geometrical optics in scalar-tensor theories.
\newblock \emph{Journal of Cosmology and Astroparticle Physics}, 2020\penalty0
  (11), 2020.

\bibitem[Garoffolo~et al.(2021)]{Garoffolo_2021}
A.~Garoffolo~et al.
\newblock Detecting dark energy fluctuations with gravitational waves.
\newblock \emph{Physical Review D}, 103\penalty0 (8), 2021.

\bibitem[Graziani~et al.(2020)]{graziani2020peculiar}
R.~Graziani~et al.
\newblock Peculiar velocity cosmology with type ia supernovae, 2020.

\bibitem[Hamaus~et a.l.(2011)]{Hamaus_2011}
N.~Hamaus~et a.l.
\newblock Optimal constraints on local primordial non-gaussianity from the
  two-point statistics of large-scale structure.
\newblock \emph{Phys. Rev. D}, 84, 2011.

\bibitem[Hamilton(1998)]{Hamilton_1998}
A.~J.~S. Hamilton.
\newblock Linear redshift distortions: A review.
\newblock \emph{The Evolving Universe}, 1998.
\newblock \doi{10.1007/978-94-011-4960-0_17}.

\bibitem[Hui and Greene(2006)]{Hui_2006}
L.~Hui and P.B. Greene.
\newblock Correlated fluctuations in luminosity distance and the importance of
  peculiar motion in supernova surveys.
\newblock \emph{Physical Review D}, 73\penalty0 (12), 2006.

\bibitem[Ilić~et al.(2021)]{euclid_2021f}
S.~Ilić~et al.
\newblock \textit{Euclid} preparation: Xv. forecasting cosmological constraints
  for the \textit{Euclid} and cmb joint analysis, 2021.

\bibitem[Jensen(2004)]{Jensen2004}
J.W. Jensen.
\newblock Supernovae light curves: An argument for a new distance modulus,
  2004.

\bibitem[Kaiser(1987)]{kaiser_1987}
N.~Kaiser.
\newblock {Clustering in real space and in redshift space}.
\newblock \emph{Monthly Notices of the Royal Astronomical Society},
  227\penalty0 (1):\penalty0 1--21, 07 1987.
\newblock ISSN 0035-8711.
\newblock \doi{10.1093/mnras/227.1.1}.

\bibitem[Khokhlov et~al.(1993)Khokhlov, Mueller, and Hoeflich]{Khokhlov1993}
A.~Khokhlov, E.~Mueller, and P.~Hoeflich.
\newblock {Light curves of type IA supernova models with different explosion
  mechanisms.}
\newblock \emph{Astronomy and Astrophysics}, 270:\penalty0 223--248, 1993.

\bibitem[Krishnan~et al.(2021)]{krishnan2021hints}
C.~Krishnan~et al.
\newblock Hints of flrw breakdown from supernovae, 2021.

\bibitem[Libanore~et al.(2021)]{Libanore_2021}
S.~Libanore~et al.
\newblock Gravitational wave mergers as tracers of large scale structures.
\newblock \emph{Journal of Cosmology and Astroparticle Physics}, 2021\penalty0
  (02):\penalty0 035–035, 2021.

\bibitem[Limber(1953)]{Limber_1953}
D.~N. Limber.
\newblock The analysis of counts of the extragalactic nebulae in terms of a
  fluctuating density field.
\newblock \emph{The Astrophysical Journal}, 117, 1953.

\bibitem[Maggiore~et al.(2020)]{maggiore2020}
M.~Maggiore~et al.
\newblock {Science case for the Einstein telescope}.
\newblock \emph{ournal of Cosmology and Astroparticle Physics}, 2020\penalty0
  (3), 2020.
\newblock \doi{10.1088/1475-7516/2020/03/050}.

\bibitem[McDonald and Seljak(2009)]{McDonald_2009}
P.~McDonald and U.~Seljak.
\newblock How to evade the sample variance limit on measurements of
  redshift-space distortions.
\newblock \emph{Journal of Cosmology and Astroparticle Physics}, 2009\penalty0
  (10), 2009.

\bibitem[Mukherjee and Wandelt(2018)]{Mukherjee_2020_lensing}
S.~Mukherjee and B.~D. Wandelt.
\newblock {Beyond the classical distance-redshift test: cross-correlating
  redshift-free standard candles and sirens with redshift surveys}.
\newblock \emph{arXiv:1808.06615}, 8 2018.

\bibitem[Mukherjee~et al.(2021)]{Mukherjee_2021}
S.~Mukherjee~et al.
\newblock Accurate precision cosmology with redshift unknown gravitational wave
  sources.
\newblock \emph{Physical Review D}, 103\penalty0 (4), 2021.

\bibitem[Namikawa(2021)]{Namikawa_2021}
T.~Namikawa.
\newblock Analyzing clustering of astrophysical gravitational-wave sources:
  luminosity-distance space distortions.
\newblock \emph{Journal of Cosmology and Astroparticle Physics}, 2021\penalty0
  (01), 2021.
\newblock ISSN 1475-7516.

\bibitem[Namikawa~et al.(2016)]{Namikawa_2016}
T.~Namikawa~et al.
\newblock Anisotropies of gravitational-wave standard sirens as a new
  cosmological probe without redshift information.
\newblock \emph{Physical Review Letters}, 116\penalty0 (12), 2016.

\bibitem[Sasaki(1987)]{Sasaki_1987}
M.~Sasaki.
\newblock The magnitude-redshift relation in a perturbed friedmann universe.
\newblock \emph{Monthly Notices of the Royal Astronomical Society}, 228, 1987.

\bibitem[Scelfo~et al.(2018)]{scelfo_2018}
G.~Scelfo~et al.
\newblock {\scshape GW×LSS}: chasing the progenitors of merging binary black
  holes.
\newblock \emph{Journal of Cosmology and Astroparticle Physics}, 2018\penalty0
  (09), 2018.

\bibitem[Scelfo~et al.(2020)]{Scelfo_2020}
G.~Scelfo~et al.
\newblock Exploring galaxies-gravitational waves cross-correlations as an
  astrophysical probe.
\newblock \emph{Journal of Cosmology and Astroparticle Physics}, 2020\penalty0
  (10), 2020.

\bibitem[Schaye~et al.(2015)]{Schaye2015}
J.~Schaye~et al.
\newblock {The EAGLE project: simulating the evolution and assembly of galaxies
  and their environments}.
\newblock \emph{Monthly Notices of the Royal Astronomical Society},
  446\penalty0 (1), 2015.
\newblock \doi{10.1093/mnras/stu2058}.

\bibitem[Seljak(2009)]{Seljak_2009}
U.~Seljak.
\newblock Extracting primordial non-gaussianity without cosmic variance.
\newblock \emph{Physical Review Letters}, 102\penalty0 (2), 2009.

\bibitem[Tinker~et al.(2008)]{Tinker2008}
J.~Tinker~et al.
\newblock Toward a halo mass function for precision cosmology: The limits of
  universality.
\newblock \emph{The Astrophysical Journal}, 688\penalty0 (2):\penalty0
  709--728, 2008.

\bibitem[Veitch~et al.(2015)]{Veitch_2015}
J.~Veitch~et al.
\newblock Parameter estimation for compact binaries with ground-based
  gravitational-wave observations using the lalinference software library.
\newblock \emph{Physical Review D}, 91\penalty0 (4), 2015.
\newblock ISSN 1550-2368.

\bibitem[Vijaykumar~et al.(2020)]{vijaykumar2020probing}
A.~Vijaykumar~et al.
\newblock Probing the large scale structure using gravitational-wave
  observations of binary black holes, 2020.

\bibitem[Viljoen et~al.(2021)Viljoen, Fonseca, and
  Maartens]{viljoen2021multiwavelength}
Jan-Albert Viljoen, José Fonseca, and Roy Maartens.
\newblock Multi-wavelength spectroscopic probes: biases from neglecting
  light-cone effects, 2021.

\bibitem[Vitale and Evans(2017)]{vitale_2017}
S.~Vitale and M.~Evans.
\newblock Parameter estimation for binary black holes with networks of
  third-generation gravitational-wave detectors.
\newblock \emph{Phys. Rev. D}, 95, 2017.

\bibitem[Witzemann~et al.(2019)]{Witzemann_2019}
A.~Witzemann~et al.
\newblock Simulated multitracer analyses with hi intensity mapping.
\newblock \emph{Monthly Notices of the Royal Astronomical Society},
  485\penalty0 (4), 2019.

\bibitem[Yang~et al.(2019)]{yang2019}
T.~Yang~et al.
\newblock Constraints on the cosmic distance duality relation with simulated
  data of gravitational waves from the einstein telescope, 2019.

\bibitem[Yoo(2009)]{Yoo_2009}
J.~Yoo.
\newblock Complete treatment of galaxy two-point statistics: Gravitational
  lensing effects and redshift-space distortions.
\newblock \emph{Physical Review D}, 79\penalty0 (2), 2009.
\newblock \doi{10.1103/physrevd.79.023517}.

\bibitem[Zhang(2018)]{zhang2018}
P.~Zhang.
\newblock The large scale structure in the 3d luminosity-distance space and its
  cosmological applications, 2018.

\bibitem[Zhao et~al.(2021)Zhao, Variu, He, Sanchez, Tamone, Chuang, Kitaura,
  Tao, Yu, Kneib, Percival, Shan, Zhao, Burtin, Dawson, Rossi, Schneider, and
  de~la Macorra]{zhao2021completed}
Cheng Zhao, Andrei Variu, Mengfan He, Daniel~Forero Sanchez, Amélie Tamone,
  Chia-Hsun Chuang, Francisco-Shu Kitaura, Charling Tao, Jiaxi Yu, Jean-Paul
  Kneib, Will~J. Percival, Huanyuan Shan, Gong-Bo Zhao, Etienne Burtin, Kyle~S.
  Dawson, Graziano Rossi, Donald~P. Schneider, and Axel de~la Macorra.
\newblock The completed sdss-iv extended baryon oscillation spectroscopic
  survey: Cosmological implications from multi-tracer bao analysis with
  galaxies and voids, 2021.

\bibitem[Zhao and Wen(2018)]{Zhao_2018}
W.~Zhao and L.~Wen.
\newblock Localization accuracy of compact binary coalescences detected by the
  third-generation gravitational-wave detectors and implication for cosmology.
\newblock \emph{Physical Review D}, 97\penalty0 (6), 2018.

\bibitem[Zheng and Weinberg(2007)]{zheng2007}
Zheng Zheng and David~H. Weinberg.
\newblock Breaking the degeneracies between cosmology and galaxy bias.
\newblock \emph{The Astrophysical Journal}, 659\penalty0 (1):\penalty0 1–28,
  Apr 2007.
\newblock ISSN 1538-4357.
\newblock \doi{10.1086/512151}.
\newblock URL \url{http://dx.doi.org/10.1086/512151}.

\bibitem[Ó~Colgáin et~al.(2021)Ó~Colgáin, Sheikh-Jabbari, and Yin]{DDE}
E.~Ó~Colgáin, M.~M. Sheikh-Jabbari, and L.~Yin.
\newblock Can dark energy be dynamical?
\newblock \emph{Physical Review D}, 104\penalty0 (2), 2021.
\newblock ISSN 2470-0029.
\newblock \doi{10.1103/physrevd.104.023510}.

\end{thebibliography}

\end{document}